\documentclass[11pt,a4paper]{article}
\pdfoutput=1 

\usepackage[table]{xcolor}
\usepackage{colortbl}
\RequirePackage{ifpdf} 
\usepackage{amsmath} 
\usepackage{mathtools}
\usepackage{cases}

\usepackage{jheppub}
\usepackage{pstricks}
\usepackage[final]{pdfpages} 
\usepackage{ifpdf}
\usepackage{textcomp} 
\usepackage{slashed}

\usepackage{color}
\usepackage{xcolor}
\definecolor{urlblue}{rgb}{0.2,0.4,0.7}
\definecolor{citegreen}{rgb}{0,0.6,0.2}
\definecolor{linkred}{rgb}{0.9,0.2,0.1}
\usepackage{hyperref}
\hypersetup{
colorlinks=true, citecolor=citegreen, linkcolor=blue, urlcolor=urlblue}

\usepackage{graphics}
\usepackage{etoolbox} 
\usepackage{fixmath}
\usepackage{psfrag}

\usepackage{notoccite} 

\usepackage{amsfonts}
\usepackage{autobreak}
\usepackage{marginnote}

\usepackage{tikz}
\usetikzlibrary{positioning,arrows}
\usetikzlibrary{decorations.pathmorphing}
\usetikzlibrary{decorations.markings}
\usetikzlibrary{shapes.geometric}
\tikzset{
    vector/.style={decorate, decoration={snake}, draw},
    provector/.style={decorate, decoration={snake,amplitude=2.5pt}, draw},
    antivector/.style={decorate, decoration={snake,amplitude=-2.5pt}, draw},
    fermion/.style={draw=black, postaction={decorate},decoration={markings,mark=at position .55 with {\arrow[draw=black]{>}}}},
    fermionbar/.style={draw=black, postaction={decorate},
                       decoration={markings,mark=at position .55 with {\arrow[draw=black]{<}}}},
    fermionnoarrow/.style={draw=black},
    gluon/.style={decorate, draw=black,decoration={coil,amplitude=4pt, segment length=5pt}},
    scalar/.style={dashed,draw=black, postaction={decorate},decoration={markings,mark=at position .55 with {\arrow[draw=black]{>}}}},
    scalarbar/.style={dashed,draw=black, postaction={decorate},decoration={markings,mark=at position .55 with {\arrow[draw=black]{<}}}},
    scalarnoarrow/.style={dashed,draw=black},
    electron/.style={draw=black, postaction={decorate},decoration={markings,mark=at position .55 with {\arrow[draw=black]{>}}}},
    bigvector/.style={decorate, decoration={snake,amplitude=4pt}, draw},
}

\newcommand{\NOdisplay}[1]{ }

\title{Polarized $q \bar{q} \rightarrow Z +$Higgs amplitudes at two loops in QCD: the interplay between vector and axial vector form
factors and a pitfall in applying a non-anticommuting $\gamma_5$}

\author{Taushif Ahmed$^{a}$, Werner Bernreuther$^{b}$, Long Chen$^{a}$ and Micha\l{} Czakon$^{b}$}
\emailAdd{taushif@mpp.mpg.de, breuther@physik.rwth-aachen.de, longchen@mpp.mpg.de, mczakon@physik.rwth-aachen.de}
\affiliation{$^a$Max-Planck-Institut f\"ur Physik, F\"ohringer Ring 6, 80805 M\"unchen, Germany\\
$^b$Institut f\"ur Theoretische Teilchenphysik und Kosmologie, RWTH Aachen University,\\ Sommerfeldstr.~16, 52056 Aachen, Germany}

\preprint{MPP-2020-37,~TTK-20-10,~P3H-20-016}

\abstract{We consider QCD corrections to two loops for the polarized amplitudes of $q{\bar q}\to Z +$ Higgs boson. 
First we show how the polarized amplitudes of $b \bar{b} \rightarrow Z h$ associated with a non-vanishing $b$-quark Yukawa coupling and a scalar or pseudoscalar Higgs boson $h$ can be built up solely from  vector form factors (FF) of properly grouped classes of diagrams, bypassing completely the need of explicitly manipulating $\gamma_5$ in dimensional regularization (up to a few ``anomalous'', i.e., triangle diagrams). 
We determine the contributions of the triangle diagrams in the heavy top limit.  
We present the analytic results of the vector FF and the triangle-diagram contributions to the axial vector FF, which are sufficient for deriving the two-loop QCD amplitudes for $b \bar{b} \rightarrow Z h$ with a CP-even and CP-odd Higgs boson $h$. 
We derive the respective Ward identity for these amplitudes, which are subsequently verified to two-loop order in QCD using these FF. 
In addition, the FF of a class of corrections to $q \bar{q} \rightarrow ZH$ proportional to the top-Yukawa coupling are obtained analytically to two-loop order in QCD in the heavy-top limit using the Higgs-gluon effective Lagrangian where the top quark is integrated out.
We address a pitfall that occurs when applying the non-anticommutating $\gamma_5$ prescription to this class of contributions that has been overlooked so far in the literature.
We attribute this issue to the fact that the absence of certain heavy-mass expanded diagrams in the infinite-mass limit of a scattering amplitude with an axial vector current depends on the particular $\gamma_5$ prescription in use.
}

\begin{document}
\allowdisplaybreaks[4]
\unitlength1cm
\maketitle
\flushbottom


\def\D{{\cal D}}
\def\DD{\overline{\cal D}}
\def\g{\overline{\cal C}}
\def\gm{\gamma}
\def\M{{\cal M}}
\def\ep{\epsilon}
\def\epm1{\frac{1}{\epsilon}}
\def\epm2{\frac{1}{\epsilon^{2}}}
\def\epm3{\frac{1}{\epsilon^{3}}}
\def\epm4{\frac{1}{\epsilon^{4}}}
\def\unM{\hat{\cal M}}
\def\ashat{\hat{a}_{s}}
\def\asmur{a_{s}^{2}(\mu_{R}^{2})}
\def\sigbar{{{\overline {\sigma}}}\left(a_{s}(\mu_{R}^{2}), L\left(\mu_{R}^{2}, m_{H}^{2}\right)\right)}
\def\sigbarn{{{{\overline \sigma}}_{n}\left(a_{s}(\mu_{R}^{2}) L\left(\mu_{R}^{2}, m_{H}^{2}\right)\right)}}
\def\unas{ \left( \frac{\hat{a}_s}{\mu_0^{\epsilon}} S_{\epsilon} \right) }
\def\rnM{{\cal M}}
\def\bt{\beta}
\def\cD{{\cal D}}
\def\cC{{\cal C}}
\def\ca{\text{\tiny C}_\text{\tiny A}}
\def\cf{\text{\tiny C}_\text{\tiny F}}
\def\ct{{\red []}}
\def\sv{\text{SV}}
\def\murOmu{\left( \frac{\mu_{R}^{2}}{\mu^{2}} \right)}
\def\bb{b{\bar{b}}}
\def\bt0{\beta_{0}}
\def\bt1{\beta_{1}}
\def\bt2{\beta_{2}}
\def\bt3{\beta_{3}}
\def\gm0{\gamma_{0}}
\def\gm1{\gamma_{1}}
\def\gm2{\gamma_{2}}
\def\gm3{\gamma_{3}}
\def\nn{\nonumber}
\def\l{\left}
\def\r{\right}
\def\T{{\cal Z}}    
\def\U{{\cal Y}}

\def\nn{\nonumber\\}
\def\ep{\epsilon}
\def\T{\mathcal{T}}
\def\V{\mathcal{V}}

\newcommand\myeq{\stackrel{\mathclap{\normalfont\mbox{\tiny FR}}}{=}}

\def\qgraf{{\fontfamily{qcr}\selectfont
QGRAF}}
\def\python{{\fontfamily{qcr}\selectfont
PYTHON}}
\def\form{{\fontfamily{qcr}\selectfont
FORM}}
\def\reduze{{\fontfamily{qcr}\selectfont
REDUZE2}}
\def\kira{{\fontfamily{qcr}\selectfont
Kira}}
\def\litered{{\fontfamily{qcr}\selectfont
LiteRed}}
\def\fire{{\fontfamily{qcr}\selectfont
FIRE5}}
\def\air{{\fontfamily{qcr}\selectfont
AIR}}
\def\mint{{\fontfamily{qcr}\selectfont
Mint}}
\def\hepforge{{\fontfamily{qcr}\selectfont
HepForge}}
\def\arXiv{{\fontfamily{qcr}\selectfont
arXiv}}
\def\Python{{\fontfamily{qcr}\selectfont
Python}}
\def\ginac{{\fontfamily{qcr}\selectfont
GiNaC}}
\def\polylogtools{{\fontfamily{qcr}\selectfont
PolyLogTools}}
\def\anci{{\fontfamily{qcr}\selectfont
Finite\_ppbk.m}}
\def\gosam{{\fontfamily{qcr}\selectfont
GoSam}}
\def\fermat{{\fontfamily{qcr}\selectfont
fermat}}
\def\xml{{\fontfamily{qcr}\selectfont
qgraf-xml-drawer}}
\def\tikz{{\fontfamily{qcr}\selectfont
TikZ}}
\def\pysecdec{{\fontfamily{qcr}\selectfont
pySecDec}}

\newcommand{\dis}{}
\newcommand{\overbar}[1]{mkern-1.5mu\overline{\mkern-1.5mu#1\mkern-1.5mu}\mkern
1.5mu}
\newcommand{\TODO}[1]{ {\color{red} #1} }

\section{Introduction}
\label{sec:intro}

A detailed investigation of the kinematical and dynamical properties of the 125 GeV Higgs boson discovered at the Large Hadron Collider (LHC), i.e., its kinematic profile as well as how it interacts with (other) known fundamental particles, remains among the major research topics of the current and future physics programs. 
In particular, the production of the Higgs boson in association with a massive electroweak vector boson, known as the VH process, plays an important role in the exploration of Higgs physics at the LHC, both for a precise study of the Higgs boson's couplings to Standard Model particles and for probing new physics. 
For instance, it supplies the main production channels behind a recent experimental triumph, the direct observation of  the Higgs boson decay to a pair of bottom quarks by the ATLAS and CMS experiments~\cite{Aaboud:2018zhk,Sirunyan:2018kst}. 
This was largely possible owing to the fact that the presence of the associated vector boson offers means to substantially reduce the Standard Model backgrounds, for instance by requiring a large transverse momentum of this vector boson~\cite{Butterworth:2008iy}.
Given foreseeable upgrades in experimental precision at future collider experiments, e.g. the high luminosity LHC program, it is very desirable to have a precise knowledge about the VH process at hadron colliders on the theoretical side as well.

In view of the aforementioned phenomenological importance of VH production, there have been many computations available in the literature on this subject aiming to improve theoretical predictions, including refs.~\cite{Brein:2003wg,Brein:2011vx,Ferrera:2014lca,Ahmed:2014cla,Li:2014bfa,Catani:2014uta,Kumar:2014uwa,Campbell:2016jau,Ferrera:2017zex,Ahmed:2019udm}. 

In ref.~\cite{Ahmed:2019udm} two of the authors of this article have presented the analytic results 
of the two-loop massless QCD corrections to the $b$-quark-induced $ZH$ process pertaining to a non-vanishing $b$-quark Yukawa coupling $\lambda_b$. 
The amplitudes for the polarized $Z$-boson states were constructed following the prescription of ref.~\cite{Chen:2019wyb}.
For the treatment of  the axial vector current vertex in dimensional regularization the prescription of refs.~\cite{Larin:1991tj,Larin:1993tq} 
was used, where the $\gamma_5$ no longer anticommutes with the Dirac matrices in D dimensions.

We consider, in this article, the QCD corrections to two loops for the polarized amplitudes of $q{\bar q}\to Z +$ Higgs boson. 
First we determine the amplitudes for  $b \bar{b} \rightarrow Z h$ at two loops associated with a non-zero $b$-quark Yukawa coupling in analytic fashion, both for a scalar $(h=H)$ and a pseudoscalar $(h=A)$ Higgs boson and for a polarized  $Z$ boson. 
We use the well-known fact that an anticommuting $\gamma_5$, denoted by $\gamma_5^{\mathrm{AC}}$ at various places of this article, can be used in D-dimensional computations~\cite{Bardeen:1972vi,Chanowitz:1979zu,Gottlieb:1979ix,Ovrut:1981ne,Espriu:1982bw,Korner:1991sx} that do not involve the Adler-Bell-Jackiw (ABJ) anomaly~\cite{Adler:1969gk,Bell:1969ts}. 
We show how the respective ``non-anomalous'' contributions that correspond to diagrams where the $Z$ boson couples to an open quark line, can be built up solely from vector form factors of properly grouped classes of diagrams whose computation does not involve the axial vector current from the outset.  Furthermore, we derive and verify explicitly the Ward identities for the QCD corrections to the   $b$-quark Yukawa coupling dependent contributions to $b \bar{b} \rightarrow Z h$, $h=H,A$, using these vector form factors. 
In addition, we determine the two-loop ``anomalous'' contributions to $b \bar{b} \rightarrow Z h$ corresponding to diagrams that involve $b$- and $t$-quark triangles and the axial vector current.

The QCD corrections to the quark-annihilation induced $q {\bar q} \to Zh$ include a class of diagrams
where the Higgs boson couples directly to a closed top-quark loop that start to appear at the two-loop order in QCD. 
As the second topic of this work, we calculate a subset of these contributions, the so-called class-I diagrams in ref.~\cite{Brein:2011vx}, to $\mathcal{O}(\alpha_s^3)$ at the amplitude level in the heavy-top limit using the Higgs-gluon effective Lagrangian of Higgs effective field theory (HEFT) where the top quark is integrated out.
Our motivation for presenting the discussion of these contributions here is that their computation within HEFT led us to uncover a pitfall in the application of a non-anticommuting $\gamma_5$ to this class of contributions: 
we found that extra counterterms are needed in addition to those that are known from the prescription of~\cite{Larin:1991tj,Larin:1993tq} in order to get correct results that obey respective Ward identities when the axial vector current is dimensionally regularized with a non-anticommuting $\gamma_5$. 
We stress that this is not to be regarded as a contradiction to the prescription of ref.~\cite{Larin:1991tj,Larin:1993tq}, because after all this is a phenomenon that is associated with the use of HEFT in the calculation of this class of diagrams.
We attribute the need for such additional counterterms to the fact that the absence of certain heavy-mass expanded diagrams in the infinite-mass limit of a scattering amplitude with an axial vector current actually depends on the particular $\gamma_5$ prescription in use.
~\\

The article is organized as follows. In section~\ref{sec:AfromV}, we consider contributions to the polarized amplitude of $b \bar{b} \rightarrow Z h$ $(h = H, A)$ that are proportional to the respective $b$-quark Yukawa coupling.
First, we determine the form factor (FF) decomposition of the amplitude of $b \bar{b} \rightarrow Z H$ 
and show how in the case of non-anomalous contributions  the axial vector FF can be obtained from the properly split FF associated with the vector current. 
The latter FF are computed to two-loop order in QCD. 
Moreover, we determine the $b$- and $t$-quark contributions to the axial FF from the ``anomalous'' diagrams with quark triangles in the limit $m_t\to\infty.$ 
Then we show how the vector and axial vector FF of the respective amplitude associated with a pseudoscalar Higgs boson can be obtained from the  FF associated  with a scalar Higgs. 
In section~\ref{sec:WI}, we derive the respective Ward identity that the amplitude of section~\ref{sec:AfromV} for 
$h=H,A$ must satisfy and check these identities using the FF computed before.      
In section~\ref{sec:qqZHinHEFT}, we consider a class of contributions, the so-called class-I contributions, to $q \bar{q} \rightarrow Z H$ proportional to the top-Yukawa coupling within HEFT where the top quark is integrated out.
First, we determine the vector FF and non-anomalous axial vector FF to two loops using an anticommuting $\gamma_5$ in dimensional regularization and show that the respective Ward identities are satisfied. 
In section~\ref{sec:G5pitfall}, we recompute the non-anomalous class-I axial FF and also the contributions from the anomalous diagrams in HEFT using a non-anticommuting $\gamma_5$ in D dimensions. 
We find that both for the non-anomalous and anomalous axial vector FF, an additional counterterm is required in order to satisfy the respective (non)anomalous Ward identity.  
Finally, we investigate the axial vector part of the non-anomalous class-I diagrams with a non-anticommuting $\gamma_5$ in the full theory without taking the limit $m_t\to\infty.$
We conclude in section~\ref{sec:conc}.

\section{Axial vector form factors from vector counterparts}
\label{sec:AfromV}

\subsection{Production of a scalar Higgs boson $H$}
\label{suse:preliminary}

 In this section, we consider first, for definiteness, the production of a scalar Higgs boson, $H$, 
in association with a massive vector boson, $Z$, through bottom quark anti-quark annihilation:
\begin{align}
\label{eq:process}
    b(p_1) + \bar{b}(p_2) \to Z(q_1) + H(q_2)\,.
\end{align}
The four-momenta of the particles in  \eqref{eq:process} 
satisfy the on-shell conditions $p_1^2=p_2^2=0,\,q_1^2=m_Z^2,\,q_2^2=m_H^2$, where $m_Z$ and $m_H$ are the masses of the $Z$ and Higgs boson, respectively. 
The Mandelstam variables are 
\begin{align}
\label{eq:Mandelstam}
    s\equiv (p_1+p_2)^2\,,~~ t\equiv (p_1-q_1)^2\, ~~\text{and}~~ u\equiv (p_2-q_1)^2 \, ,
\end{align}
satisfying $s+t+u=q_1^2+q_2^2=m_Z^2+m_H^2$.

\begin{figure}[htbp]
\begin{center}
\includegraphics[scale=0.38]{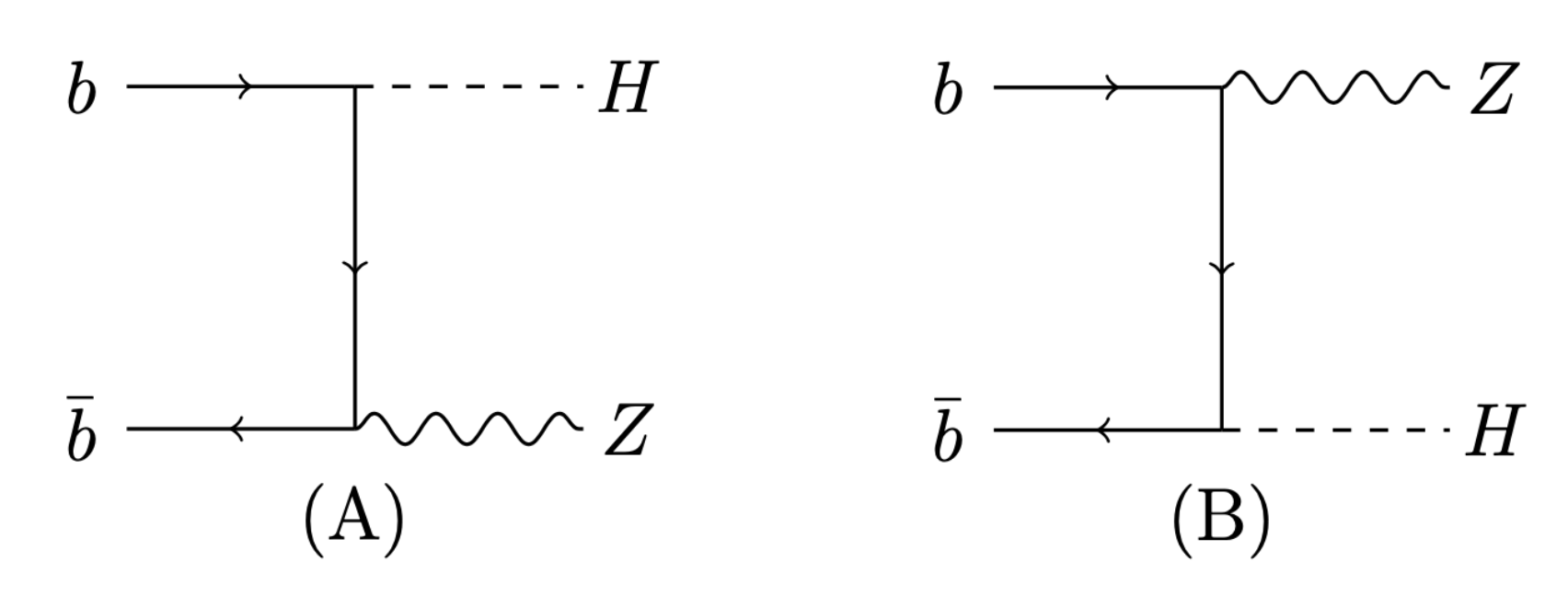}
\caption{Leading order Feynman diagrams that involve the bottom-Higgs Yukawa coupling.}
\label{dia:tree}
\end{center}
\end{figure}

We keep a non-zero Yukawa coupling only for the $b$ quark, 
which otherwise is taken to be massless, and for the top quark whose contributions are considered in the infinite mass limit. 
By employing an elegant methodology, we recompute in this section the two-loop QCD corrections to the non-Drell-Yan type diagrams of the process \eqref{eq:process}, shown at tree level in figure~\ref{dia:tree},
that depend on the $b$-quark Yukawa coupling $\lambda_b$, in QCD for 5 massless flavors and determine also the $t$-quark loop-induced contributions in the infinite mass limit. 
These contributions form a gauge-invariant set.
The amplitude of these contributions can be parameterized, including QCD corrections to two loops,  by 
\begin{align} 
\label{eq:bbZHamplitude}
    \mathcal{M} &= \lambda_b\, \bar{v}(p_2) \, \mathbf{\Gamma}^{\mu} \, u(p_1) \, \varepsilon^{*}_{\mu}(q_1) \nonumber\\
    &=\lambda_b~ g_{V,b}  \bar{v}(p_2) \, \mathbf{\Gamma}^{\mu}_{vec} \, u(p_1) \, \varepsilon^{*}_{\mu}(q_1) 
    \,+\,\lambda_b~ g_{A,b} \bar{v}(p_2) \, \mathbf{\Gamma}^{\mu}_{axi} \, u(p_1) \, \varepsilon^{*}_{\mu}(q_1)\nonumber\\
    &\equiv  \lambda_b~ g_{V,b} {\cal M}_{vec}+ \lambda_b~ g_{A,b} {\cal M}_{axi}\,.
\end{align}
The symbol $\mathbf{\Gamma}^{\mu} \equiv  g_{V,b}\mathbf{\Gamma}^{\mu}_{vec} +  g_{A,b}\mathbf{\Gamma}^{\mu}_{axi}$ represents 
a matrix in the spinor space with one open Lorentz index $\mu$ that may be carried by either the  Dirac matrix $\gamma^{\mu}$ or 
one of the external momenta involved. 
And it is the sum of the contributions from the vector and axial vector couplings of the $Z$ boson. 
For purposes discussed below we have factored out $\lambda_b$ and the vector and axial vector couplings of the $b$ quark to the $Z$ boson, denoted by $g_{V,b}$ and $g_{A,b}$, respectively.

At the end of this section, we consider the production of a pseudoscalar Higgs boson $A$ analogous to \eqref{eq:process} and discuss how the respective scattering amplitude to two-loop order in QCD analogous to \eqref{eq:bbZHamplitude} can be obtained from the vector form factors that determine the amplitude \eqref{eq:bbZHamplitude} and which will be computed next.

\subsection{The interplay between axial vector and vector form factors}
\label{suse:intva}

\begin{figure}[htbp]
\begin{center}
\includegraphics[scale=0.65]{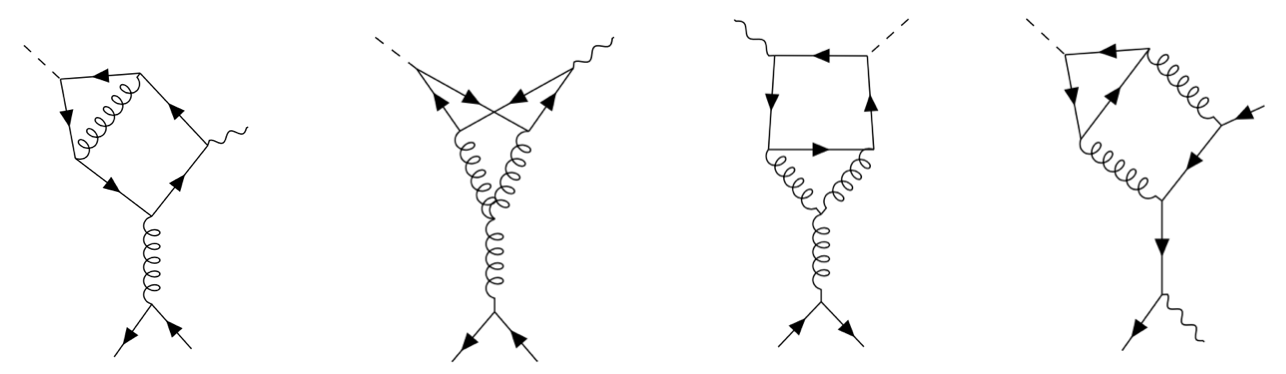}
\caption{Examples of diagrams where the Higgs boson or the Higgs and the $Z$ boson are coupled to a closed quark loop.}
\label{dia:H-closed-fermion}
\end{center}
\end{figure}

Unlike ref.~\cite{Ahmed:2019udm}, where a non-anticommuting $\gamma_5$ was used for the $Z$ boson axial vector coupling, we compute here all non-anomalous contributions to the amplitude \eqref{eq:bbZHamplitude} to two loops using an anticommuting $\gamma_5^{\rm AC}$ in $D$ dimensions. First we consider all two-loop diagrams where i) only the Higgs boson and ii) the Higgs and $Z$ boson are radiated from a closed massless quark loop, such as those shown in figure~\ref{dia:H-closed-fermion}.
These contributions vanish because they involve a trace with an odd number of Dirac $\gamma$ matrices. 
Thus, in all non-vanishing non-anomalous contributions to two-loop order in QCD, the $Z$ boson couples to the open $b$-quark line and $\gamma_5^{\rm AC}$ contained in $\mathbf{\Gamma}^{\mu}_{axi}$  can be anticommuted next to an external $b$-quark spinor.

These non-anomalous diagrams can be further divided into two classes, denoted in the following by  class-$ZH$ and class-$HZ$, 
which correspond to the QCD corrections 
to the tree-level diagrams (A) and (B), respectively, of figure~\ref{dia:tree}.
The reason for this separation  is simply the fact that due to the presence of the chirality-flipping Yukawa 
interaction on the $b$-quark line  a relative minus sign is generated between these two contributions when the anticommuting $\gamma_5^{AC}$ 
is pushed next to the same external $b$-quark spinor.
This is also the reason why here the non-anomalous axial vector form factors are not identical to the vector ones, see below.
~\\

Let us turn off for a moment the axial vector coupling of the $Z$ boson and consider only ${\cal M}_{vec}$ in \eqref{eq:bbZHamplitude}.
We can decompose it in terms of form factors as follows:
\begin{align} 
\label{eq:bbZHamplitude_VECS}
    \mathcal{M}_{vec} &= 
    \bar{v}(p_2) \, \mathbf{\Gamma}^{\mu}_{ZH} \, u(p_1) \, \varepsilon^{*}_{\mu}(q_1) 
    \,+\,\bar{v}(p_2) \, \mathbf{\Gamma}^{\mu}_{HZ} \, u(p_1) \, \varepsilon^{*}_{\mu}(q_1)\,, \nonumber\\
   \bar{v}(p_2) \, \mathbf{\Gamma}^{\mu}_{X} \, u(p_1) &= 
    F_{1,X} \, \bar{v}(p_2) \, u(p_1) \, p_1^{\mu} 
    \,+\, F_{2,X} \, \bar{v}(p_2) \, u(p_1) \, p_2^{\mu} \nonumber\\
    &\,+\, F_{3,X} \, \bar{v}(p_2) \, u(p_1) \, q_1^{\mu}
    \,+\, F_{4,X} \, \bar{v}(p_2) \,\gamma^{\mu} \slashed{q}_1 u(p_1) \,, \quad X=ZH, HZ. 
\end{align}
In this form factor decomposition, which reflects the chirality flip along the massless $b$-quark line,
we have taken into account  the equations of motion for the on-shell massless spinors $\bar{v}(p_2)$ and $u(p_1)$, 
 but have not confined ourselves to the physical polarization states of the $Z$ boson. 
 The projectors for obtaining  the 
vector form factors defined in \eqref{eq:bbZHamplitude_VECS}  from $\mathbf{\Gamma}^{\mu}_{X}$ $(X=ZH, HZ)$ are derived and explicitly given in ref.~\cite{Ahmed:2019udm}.
In the absence of the axial current, the four basis structures in \eqref{eq:bbZHamplitude_VECS} 
are linearly complete in D dimensions for $\mathcal{M}_{vec}$ in \eqref{eq:bbZHamplitude}, 
 regardless of the QCD loop order.

The projection of the two sets of vector form factors $F_{i,ZH}$ and $F_{i,HZ}$ encounters no subtlety at all, 
and their renormalization is standard with the details given in ref.~\cite{Ahmed:2019udm}.
The complete vector form factors, $F_{i,vec}$, defined 
by  
\begin{align}
\label{eq:complete-vec-FF}
    {\cal M}_{vec}^\mu &= F_{1,vec} \, \bar{v}(p_2) \, u(p_1) \, p_1^{\mu} 
    \,+\, F_{2,vec} \, \bar{v}(p_2) \, u(p_1) \, p_2^{\mu} \nonumber\\
    &\,+\, F_{3,vec} \, \bar{v}(p_2) \, u(p_1) \, q_1^{\mu}
    \,+\, F_{4,vec} \, \bar{v}(p_2) \,\gamma^{\mu} \slashed{q}_1 u(p_1)\,,
\end{align}
are given by
\begin{align} 
\label{eq:bbZHamplitude_vecs}
F_{i,vec} =  F_{i,ZH} + F_{i,HZ} \quad \text{for} \; i = 1,2,3,4 \,.
\end{align}
 Restoring the axial vector couplings of the $Z$ boson, the  amplitude ${\cal M}_{axi}$ 
  defined in eq.~\eqref{eq:bbZHamplitude} consists of a ``non-anomalous'' and  ``anomalous''
  contribution:
  \begin{equation}
   \label{eq:axisns}
   {\cal M}_{axi} = {\cal M}_{axi(ns)} + {\cal M}_{axi(s)} \, .
  \end{equation}
These contributions can be decomposed into form factors, in analogy to eq.~\eqref{eq:complete-vec-FF}.
Using  $\gamma_5^{\rm AC}$ in D dimensions,
 the non-anomalous axial form factors $F_{i,axi(ns)}$, defined by
\begin{align}
\label{eq:complete-axi-FF}
\mathcal{M}_{axi(ns)}^{\mu} &= 
    F_{1,axi(ns)} \, \bar{v}(p_2) \,\gamma_5\, u(p_1) \, p_1^{\mu}
    \,+\, F_{2,axi(ns)} \, \bar{v}(p_2) \,\gamma_5\, u(p_1) \, p_2^{\mu} \nonumber\\
    &\,+\, F_{3,axi(ns)} \, \bar{v}(p_2) \,\gamma_5\, u(p_1) \, q_1^{\mu}
    \,+\, F_{4,axi(ns)} \, \bar{v}(p_2) \,\gamma^{\mu} \slashed{q}_1 \gamma_5 u(p_1) \, ,
\end{align}
are obtained by 
\begin{align} 
\label{eq:bbZHamplitude_axis}
F_{i,axi(ns)} = F_{i,HZ} - F_{i,ZH}\,.
\end{align}
The appearance of the relative minus sign in \eqref{eq:bbZHamplitude_axis} is explained above eq.~\eqref{eq:bbZHamplitude_VECS}.

We have checked that the unrenormalized, unsubtracted vector form factors  $F_{i,vec}$ 
 defined in \eqref{eq:bbZHamplitude_vecs} agree with those obtained in ref.~\cite{Ahmed:2019udm}
to two-loop order in D dimensions. Without surprise, the axial form factors $F_{i,axi}$ composed in  \eqref{eq:bbZHamplitude_axis} are 
indeed different from those defined in ref.~\cite{Ahmed:2019udm} in their bare form. 
However, after carrying out the ultraviolet (UV) renormalizations and infrared (IR) subtractions, 
the finite remainders of these form factors are identical to those given in ref.~\cite{Ahmed:2019udm}.\footnote{There is a relative minus sign between
the fourth axial form factor defined by \eqref{eq:bbZHamplitude_axis} and the corresponding one defined by equation (3.19) in ref.~\cite{Ahmed:2019udm}, 
because the latter one  corresponds to the Lorentz structure $\bar{v}(p_2) \,\gamma^{\mu} \gamma_5  \slashed{q}_1  u(p_1)$.} 
Here we remark that the UV renormalization needed for $F_{i,axi(ns)}$ in \eqref{eq:bbZHamplitude_axis} 
is identical to that of the vector form factors, which is different from what was done in ref.~\cite{Ahmed:2019udm} regarding the axial form factors.
We have thus cross-checked the previous computation of the non-anomalous axial part of ref.~\cite{Ahmed:2019udm}
(where a non-anticommuting $\gamma_5$~\cite{Larin:1991tj} was used). 
The UV renormalized results of the partial vector form factors, $F_{i,ZH}$ and $F_{i,HZ}$, which are the building blocks for composing the complete non-anomalous vector and axial matrix elements, are provided as provided as supplementary material.

Details about conventions and variables of the analytic expressions can be found in the \textbf{{\small ReadMe.txt}} submitted as supplementary material.
We present the results to two-loop order, more specifically, the first three coefficients of the partial FF in the expansion
\begin{align}
\label{eq:partial-FF-expansion}
    F_{i,X}=\lambda_b(\mu_R^2) \sum_{l=0}^{\infty} a_s^l(\mu_R^2) F_{i,X}^{(l)}\,,
\end{align}
where $X \in \{ZH,HZ\}$, $a_s=\alpha_s/(4\pi)$ and $\mu_R$ is the renormalization scale.

\subsection{Contributions from diagrams involving quark triangles}
\label{suse:anomalousdiagrams}

\begin{figure}[htbp]
\begin{center}
\includegraphics[scale=0.4]{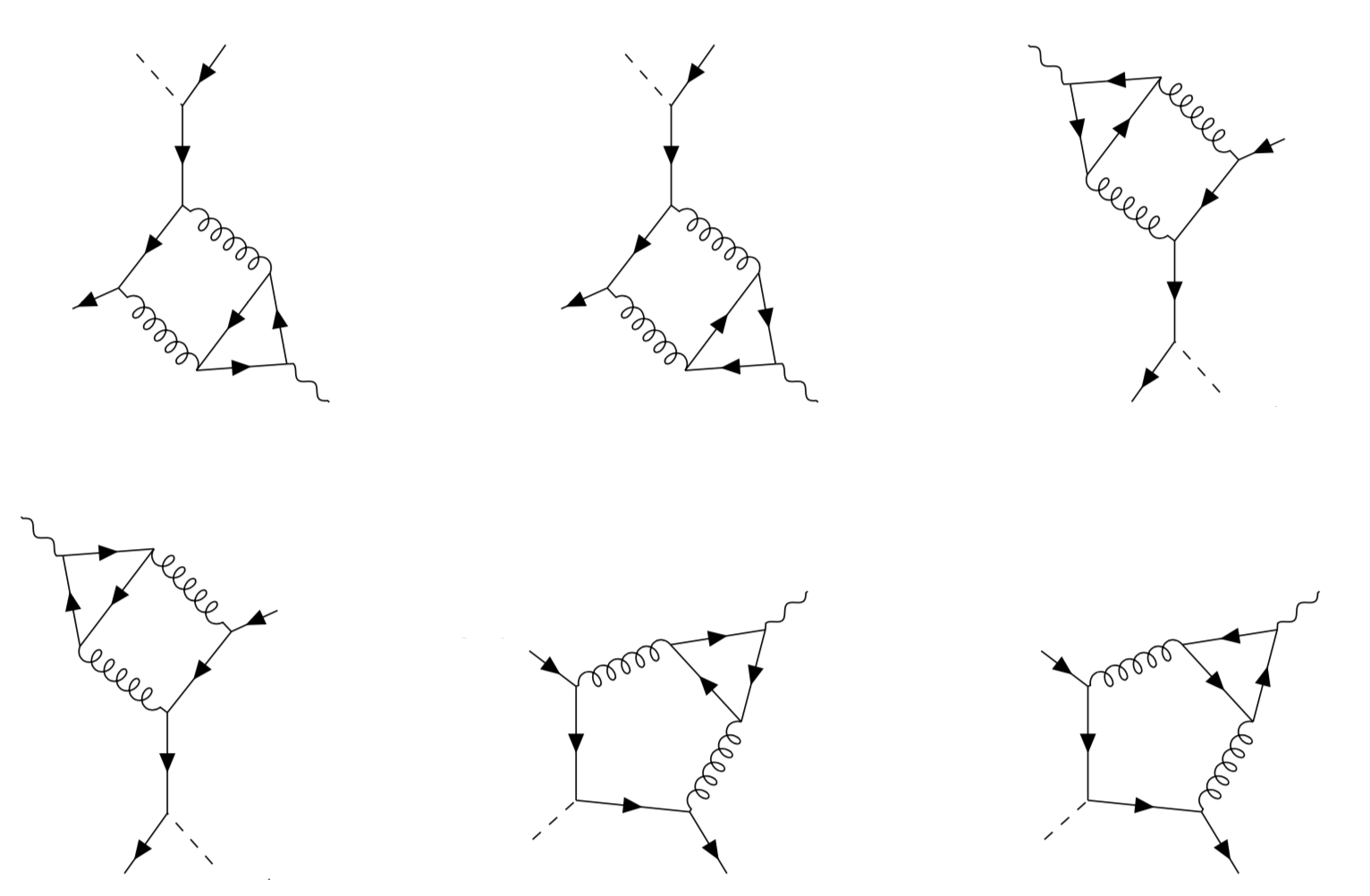}
\caption{The two-loop diagrams proportional to $\lambda_b$ that involve quark triangles.
 We denote the first four diagrams by box-triangles and the last two by pentagon-triangles.}
\label{dia:singlet}
\end{center}
\end{figure}

The two-loop ``anomalous'' contributions to the process \eqref{eq:process}  involving quark triangles can be  represented by six Feynman diagrams, 
shown in figure~\ref{dia:singlet}.
Furry's theorem tells us that only the axial vector part is non-vanishing. 
The contributions of the massless $u$- and $d$-type quarks circulating in the fermion triangle of figure~\ref{dia:singlet} cancel; 
thus only the mass non-degenerate $b$ and $t$ quark contribute.
Therefore we have 
\begin{equation}
   \label{eq:axisnst}
    {\cal M}^\mu_{axi(s)}  = {\cal M}^{b,\mu}_{axi(s)}  - {\cal M}^{t,\mu}_{axi(s)}(m_t) \, 
\end{equation}
where $m_t$ is the top quark mass. The relative minus sign is due to the definition of ${\cal M}_{axi}$ in eq.~\eqref{eq:bbZHamplitude} 
(where we pulled out an overall coupling factor $g_{A,b}$ from the contribution of the axial vector) and the opposite weak isospins of the $b$ and $t$ quark. 
The amplitude ${\cal M}_{axi(s)}^\mu$ can be decomposed into form factors $F_{i,axi(s)}$, in complete analogy to eq.~\eqref{eq:complete-axi-FF}.

The top-loop induced diagrams that contribute to the process \eqref{eq:process}, all starting at $\mathcal{O}(\alpha^2_s)$, 
can be divided into a set which is independent of $t$-quark Yukawa coupling $\lambda_t$, i.e.~the diagrams of figure~\ref{dia:singlet} with the $t$ quark 
circulating in the triangle loop, 
and another set dependent on the top Yukawa coupling that will be addressed in section \ref{sec:qqZHinHEFT}.
Concerning the contributions to ${\cal M}^{t,\mu}_{axi(s)}(m_t)$, we show below that 
the last two pentagon-triangle diagrams of figure~\ref{dia:singlet} vanish in the limit $m_t \rightarrow \infty$. 
The contributions of the other four box-triangle diagrams are not zero in this limit,
but they can be captured  by a decoupling  matching constant introduced for the effective $b\bar{b}Z$ interaction,
resulting from integrating out the virtual top quark in these diagrams.

\begin{figure}[htbp]
\begin{center}
\includegraphics[scale=0.35]{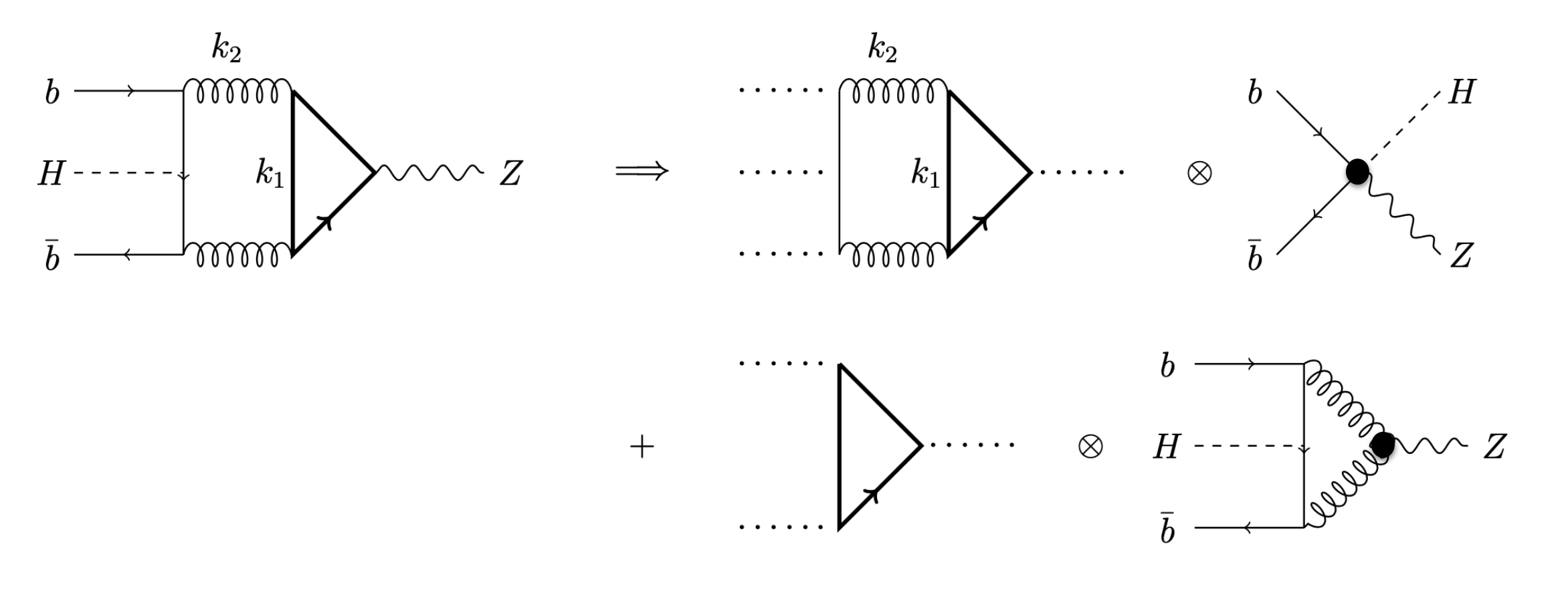}
\caption{
The heavy-mass expansion of a representative pentagon-triangle diagram with $k_{1,2}$ denoting the two loop momenta.  
The solid thick lines represent the massive $t$-quark propagators. 
The dotted lines in the asymptotic irreducible graphs~\cite{Smirnov:2002pj} on the left of ``$\otimes$'' indicate external  momenta which can all be set to zero 
if only the leading contribution in the limit  $m_t \rightarrow \infty$ is to be kept. 
The graphs on the right of ``$\otimes$'' denote the corresponding co-graphs in the heavy-mass expanded result.
}
\label{dia:mtinfinity1}
\end{center}
\end{figure}

Applying the expansion-by-graph procedure (see, e.g.~ref.\cite{Smirnov:2002pj} and references therein) 
 to one representative pentagon-triangle  diagram, as shown on the l.h.s. of figure \ref{dia:mtinfinity1}, 
 in the heavy-top limit $m_t \rightarrow \infty$, one ends up with the sum of two classes of heavy-mass expanded contributions, 
 indicated by the two terms on the r.h.s. of this figure. 
They correspond to the \textit{hard-hard} loop momentum region with $|k_2| \sim |k_1| \sim m_t$ and the \textit{soft-hard} loop momentum 
region with\footnote{The regions where $k_1$ is small do not contribute here because they lead to \textit{scaleless} loop integrals 
from the massive triangle subgraph.} $|k_2| < |k_1| \sim m_t$, respectively.
Let us first look at the contribution from the \textit{hard-hard} momentum region, corresponding to the first term 
on the r.h.s. of figure \ref{dia:mtinfinity1}.
The asymptotically irreducible graph~\cite{Smirnov:2002pj} here is a two-loop ``vacuum'' graph that has the mass-dimension -1, 
and after expansion in small ratios it depends polynomially on the external momenta which can be set to zero as $m_t$ goes to infinity and hence only on $m_t$ in the heavy-top limit. 
This is sufficient to ensure that there is no non-vanishing contribution from this graph in the limit $m_t \rightarrow \infty$.

\begin{figure}[htbp]
\begin{center}
\includegraphics[scale=0.35]{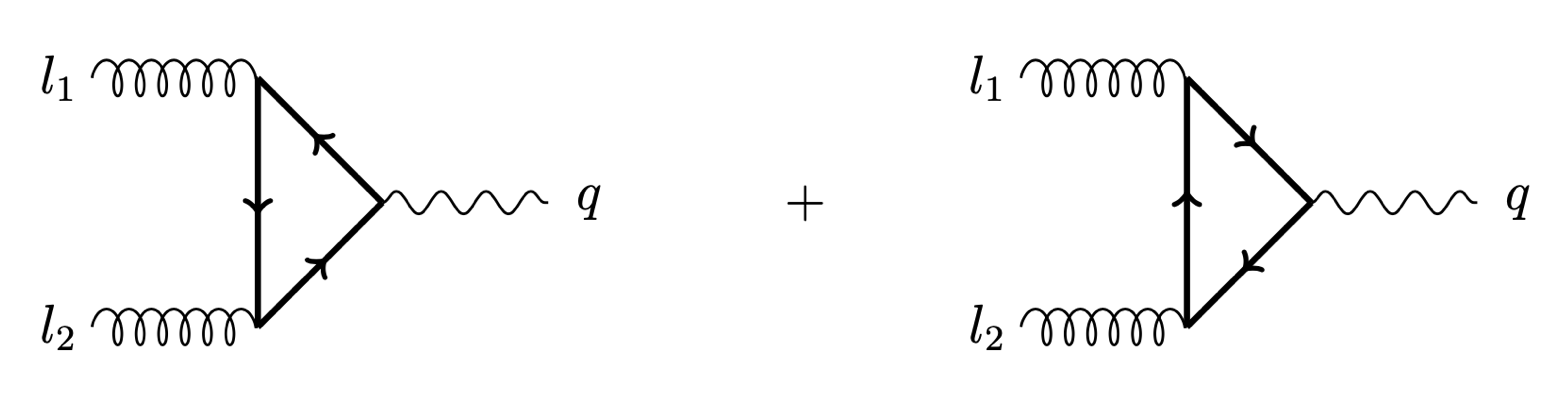}
\caption{
The VVA triangle diagrams appearing as subgraphs in the heavy-mass expansion of figure \ref{dia:mtinfinity1}.
}
\label{dia:vva}
\end{center}
\end{figure}

We then move on to the second term, originating from the \textit{soft-hard} momentum region. 
Here the subgraph that is to be expanded is a triangle $t$-quark loop that has the same topology as the VVA triangle diagrams  drawn in figure \ref{dia:vva}.
Here, counting of the mass dimension alone is no longer sufficient to tell us whether or not this subgraph has a non-vanishing limit when $m_t$ goes to infinity. 
Therefore we compute this subgraph  explicitly and take the analytic expressions of all one-loop integrals involved from their implementations in PackageX~\cite{Patel:2015tea}.

Leaving all Lorentz indices from the three gauge vertices open in the 
diagrams of figure \ref{dia:vva}, a rank-3 Lorentz tensor amplitude $\Gamma^{\mu_1  \mu_2 \mu}(l_1, l_2, m_t)$ is thus introduced for the sum of these two one-loop diagrams. 
The tensor $\Gamma^{\mu_1  \mu_2 \mu}(l_1, l_2, m_t)$ can be further split
into a vector (spin-1) part and a scalar (spin-0) part with respect to its axial-vector current index $\mu$: 
\begin{align}
\label{eq:VVAtensoramp}
   \Gamma^{\mu_1  \mu_2 \mu}(l_1, l_2, m_t) &= \Gamma^{ \mu_1  \mu_2 \nu}(l_1, l_2, m_t)\, \left(g^{\mu}_{~\nu} - \frac{q^{\mu}q_{\nu}}{q^2}\right) 
   \,+\, \Gamma^{ \mu_1  \mu_2 \nu}(l_1, l_2, m_t)\, \frac{q^{\mu}q_{\nu}}{q^2}\,, \nonumber\\
   &\equiv -\Gamma^{\mu_1  \mu_2 \mu}_v(l_1, l_2, m_t) \,+\, \Gamma^{\mu_1  \mu_2 \, q}_s(l_1, l_2, m_t)\, \frac{q^{\mu}}{q^2}\, ,
\end{align}
where $q \equiv l_1+l_2\,$, $\Gamma^{\mu_1  \mu_2 \, q}_s(l_1, l_2, m_t) \equiv \Gamma^{ \mu_1  \mu_2 \nu}(l_1, l_2, m_t) \, q_{\nu}$, and the physical polarization projector $g_{\mu\nu} - \frac{q_{\mu}q_{\nu}}{q^2}$ projects $\Gamma^{\mu_1  \mu_2 \mu}(l_1, l_2, m_t)$ onto the space of vector polarizations (indicated by the subscript $v$). 
We refrain from going into the technical details of the one-loop computation involved, e.g.~via the form factor decomposition approach, but merely point out the following explicitly verified fact: keeping the momenta $l_1$ and $l_2$ fixed, the rank-3 Lorentz tensor amplitude $\Gamma^{\mu_1  \mu_2 \mu}(l_1, l_2, m_t)$ vanishes in the limit $m_t \rightarrow \infty$.
This holds true for $\Gamma^{\mu_1  \mu_2 \mu}_v(l_1, l_2, m_t)$ and $\Gamma^{\mu_1  \mu_2 \, q}_s(l_1, l_2, m_t)$, respectively.
This implies that the second term on the r.h.s. of figure \ref{dia:mtinfinity1} associated 
with the \textit{soft-hard} region vanishes in the limit $m_t \rightarrow \infty$. 
Therefore the two pentagon-triangle diagrams do not contribute in this limit.~\\

As an aside we briefly comment on an interesting implication related to the above statement about the triangular one-loop subgraph, which, albeit not new, does not seem to be common knowledge. 
At the on-shell kinematic configuration $l_1^2 = l_2^2 = 0$, the term $\Gamma^{\mu_1  \mu_2 \mu}_v(l_1, l_2, m_t)$ vanishes completely in four dimensions, owing to the 
Landau-Yang theorem (because the color factor here is trivial), regardless of the mass of the top quark. 
The only non-vanishing piece at this configuration is $\Gamma^{\mu_1  \mu_2 \, q}_s(l_1, l_2, m_t)$, 
associated with the scalar polarization state. 
For $m_t =0$, $\Gamma^{\mu_1  \mu_2 \, q}_s(l_1, l_2, m_t)$ is given precisely by the ABJ anomaly.
What we would like to emphasize here is that $\Gamma^{\mu_1  \mu_2 \, q}_s(l_1, l_2, m_t)$ from the one-loop diagrams of figure \ref{dia:vva} vanishes in the heavy-top limit $m_t \rightarrow \infty$, as a consequence of a non-trivial cancellation between the pure $m_t$-independent quantum anomalous contribution (i.e.~the ABJ anomaly) and the non-vanishing limit of the $m_t$-dependent ``classical'' contribution at $m_t \rightarrow \infty$.
In other words, the non-vanishing $m_t \rightarrow \infty$ limit of the non-decoupling $m_t$-dependent ``classical'' contribution to $\Gamma^{\mu_1  \mu_2 \, q}_s(l_1, l_2, m_t)$ is exactly opposite to the $m_t$-independent ABJ anomaly contribution. 
Indeed, this point can be checked straightforwardly by a direct projection, because $\Gamma^{\mu_1  \mu_2 \, q}_s(l_1, l_2, m_t)$ contains just one Lorentz covariant structure. 
The corresponding unique projector is indifferent to whether the contribution comes from the
$m_t$-independent quantum anomalous part or the $m_t$-dependent ``classical'' part, and both will be projected out on the same footing.
~\\

\begin{figure}[htbp]
\begin{center}
\includegraphics[scale=0.35]{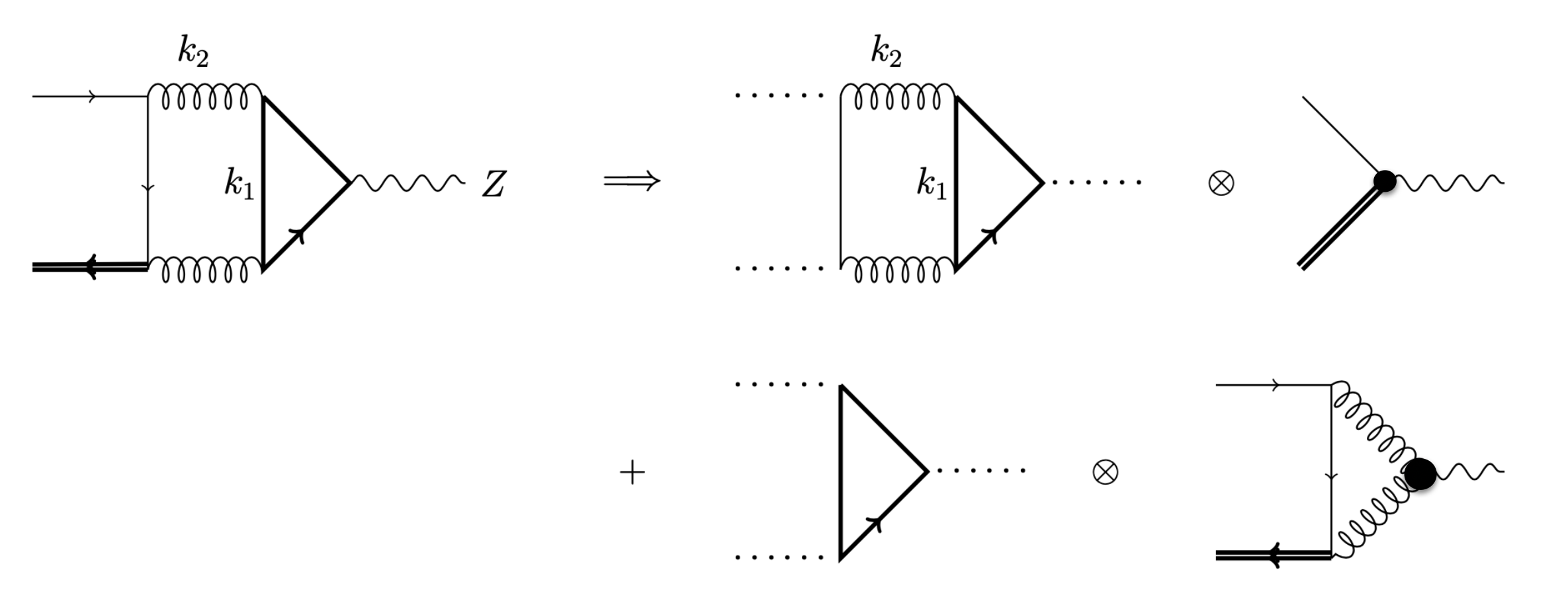}
\caption{
The heavy-mass expansion of a representative box-triangle  diagram with $k_{1,2}$ denoting the two loop momenta,
drawn in a similar fashion as in figure~\ref{dia:vva}, albeit, with a difference:
 the double-thick line represents the amputated $b$-quark tree propagator, not to be confused with the solid thick triangle representing top quark loop.
}
\label{dia:mtinfinity2}
\end{center}
\end{figure}

Let us come back to the discussion of the heavy-top limit of the remaining four box-triangle diagrams of figure~\ref{dia:singlet} where the results about the triangle subgraphs given above will be used again. 
Applying the expansion-by-graph procedure to one representative diagram of this topology, as shown on the l.h.s. of figure \ref{dia:mtinfinity2}, in the heavy-top limit, one ends up with the sum of two classes of heavy-mass expanded contributions, similar to the case of the pentagon-triangle diagrams. 
Note that here the $b$-quark tree propagator is amputated before applying the heavy-mass expansion procedure, 
resulting in an off-shell $b$-quark leg (indicated by the double-thick line in figure \ref{dia:mtinfinity2}) whose momentum 
is considered small compared to $m_t$.
The second term on the r.h.s. of figure \ref{dia:mtinfinity2} associated with the \textit{soft-hard} loop-momentum 
region vanishes in the limit of infinitely heavy top quark due to the same reason as mentioned above. 
However, the first term that originates from the \textit{hard-hard} momentum region has an asymptotic irreducible graph 
that is a two-loop box-triangle ``vacuum'' graph with zero mass-dimension and turns out to have a non-vanishing limit at $m_t \rightarrow \infty$. 
As should be clear from the diagrammatic illustration of the heavy-mass expansion result given in figure~\ref{dia:mtinfinity2}, 
this contribution can be presented 
 by a set of local composite operators determined by the co-graph, an effective $bbZ$-vertex, and the ``vacuum'' graph in front of it.
One key feature of the expanded ``vacuum'' graph (associated with the \textit{hard-hard} momentum region) 
 is that its dependence on the external momenta $\{p_1, p_2,q_1, q_2\}$ is purely polynomial.  
In other words, this special vacuum loop amplitude has a regular limit for vanishing external momenta.
Since we are only interested in the leading contribution in $1/m_t$ at $m_t \rightarrow \infty$ 
 where all external momenta in this expanded ``vacuum'' graph can be put to zero, it is then not hard to see that this expanded box-triangle ``vacuum'' graph will lead to the same expression regardless of whether or not the $b$-quark line is on-shell. 
This non-vanishing infinite-$m_t$ limit of the (properly renormalized) r.h.s.
of figure \ref{dia:mtinfinity2} can thus be captured by introducing an effective $b\bar{b}Z$ interaction with a decoupling matching constant $C_{bbZ}^{(s)}$ (being independent of the kinematics), denoted by   
\begin{center}
$C_{bbZ}^{(s)}(m_t) \, \bar{b} \,\gamma^{\mu} \gamma_5 \, b Z_{\mu}$, 
\end{center} 
accounting for integrating out the $t$-triangle loop from these diagrams at the leading power in the heavy-top expansion.\footnote{Note that there will be additional effective  operators involving the space-time derivative acting on fields starting from sub-leading power corrections in the $1/m_t$ expansion.}
Based on the information above, $C_{bbZ}^{(s)}(m_t)$ can be extracted from the heavy-top limit of the two-loop anomalous QCD corrections to the renormalized (on-shell) quark form factors computed in ref.~\cite{Bernreuther:2005rw}, and is given by
\begin{equation}
\label{eq:bbZoperator}
C_{bbZ}^{(s)}(m_t) = g_{A,\,t}\,a^2_s \, 4 C_F T_R\left( \frac{3}{2} - 3\,\mathrm{ln}\left( \frac{\mu_R^2}{m^2_t}\right) \right) \,
\end{equation} 
where $g_{A,t}$ denotes the axial coupling between the top quark and $Z$ boson, $C_F=(N_c^2-1)/(2 N_c)$ and $T_R=1/2$.  
We note in passing that this expression is equal to twice the order-$a_s^2$ Wilson coefficient in front of the effective 
interaction $q\bar{q}A$ between a pair of light quarks and a pseudoscalar Higgs boson $A$ resulting from integrating out the heavy top quark, computed in ref.~\cite{Chetyrkin:1998mw}.

In summary, the non-vanishing infinite-$m_t$ limit of ${\cal M}^t_{axi(s)}(m_t)$ of the diagrams of
figure~\ref{dia:singlet} with a $t$-quark triangle is given by
 the axial part of the tree-level amplitude of the process \eqref{eq:process} multiplied by the coefficient \eqref{eq:bbZoperator}.
We attach the analytic results of the properly renormalized axial form factors of the complete 
  contribution ${\cal M}_{axi(s)}$ of \eqref{eq:axisnst}, defined in complete analogy to \eqref{eq:complete-axi-FF}, as ancillary files 
  in Mathematica format along with the arXiv submission. (See ref.~\cite{Ahmed:2019udm} for the renormalization constants used.) 
Finally, we remark that after incorporating the non-vanishing contribution ${\cal M}^t_{axi(s)}(m_t)$, the explicit $\mu_R$ dependence 
cancels in the amplitude ${\cal M}_{axi(s)} = {\cal M}^b_{axi(s)} - {\cal M}^t_{axi(s)}(m_t)$ at the two-loop order.

\subsection{Production of a pseudoscalar Higgs boson $A$}
\label{suse:bbZAamp}

 Now let us discuss the production of a CP-odd Higgs boson $A$ in association with a $Z$ boson by $b{\bar b}$ annihilation,
\begin{align}
\label{eq:Aprocess}
    b(p_1) + \bar{b}(p_2) \to Z(q_1) + A(q_2)\,.
\end{align}
 We remark that in this case the two diagrams of figure~\ref{dia:tree} represent the complete tree-level amplitude (because a tree-level $ZZA$ vertex does not exist). 
 Using $\gamma_5^{\rm AC}$ in D dimensions the  amplitude 
\begin{equation}
 \label{eq:Aloamp}
 {\cal M}^A ={\tilde\lambda}_b~ g_{V,b} {\cal M}^A_{vec} +{\tilde\lambda}_b~ g_{A,b} {\cal M}^A_{axi}
\end{equation}
 proportional to the pseudoscalar $b$-quark Yukawa coupling ${\tilde\lambda}_b$ can be constructed to two-loop order QCD with the form factors computed above. 
Notice that the subscripts, $vec$ and $axi$, of the amplitudes in \eqref{eq:Aloamp} refer to the respective coupling between the $Z$ boson and the $b$ quark. 
It is straightforward to see that the set of Lorentz basis  structures, and consequently the corresponding form factor projectors previously used, 
in the form factor decomposition for the amplitude involving the CP-even Higgs boson $H$ exchange their roles in the current process \eqref{eq:Aprocess}. 
For instance, the set of Lorentz structures for decomposing  ${\cal M}_{vec}$ appearing in \eqref{eq:complete-vec-FF} is now the one needed for the form factor decomposition of ${\cal M}^A_{axi}$ defined in \eqref{eq:Aloamp}. 
Consequently the corresponding ``vector'' projectors  
will project out the axial form factors of the production amplitude of a pseudoscalar Higgs boson $A$ in association with a $Z$ boson.

It can be verified to two-loop order, conveniently using  $\gamma_5^{\rm AC}$, that the following relationships hold among the form factors:
\begin{align}
\label{eq:fofacA}
& F^A_{i,vec} = F_{i,vec} = F_{i,HZ} + F_{i,ZH}\,, \quad \nonumber\\ 
& F^A_{i,axi(ns)} = F_{i,axi(ns)} = F_{i,HZ} - F_{i,ZH} \, , \quad i=1,2,3,4, 
\end{align}
(up to an overall phase factor depending on the parameterization of the general Yukawa couplings. 
For our choice \eqref{eq:lagbhz} it is an overall factor $i$. It is suppressed here for simplicity.). 
The results of $F_{i,ZH}$ and $F_{i,HZ}$ were obtained in section \ref{suse:intva}.  
Furthermore, for the contribution from the triangle diagrams analogous to figure~\ref{dia:singlet}, we have 
\begin{equation}
\label{eq:fofacAs}
F^A_{i,axi(s)} = F_{i,axi(s)} \, , \quad i=1,2,3,4,
\end{equation}
 and the computation of $F_{i,axi(s)}$ of ${\cal M}_{axi}$ was discussed in section \ref{suse:anomalousdiagrams}.

\section{The Ward identities for $b \bar{b} \rightarrow ZH,~Z A$}
\label{sec:WI}

In this section we  derive and subsequently check, using the form factors $F_{i,HZ}$ and $F_{i,HZ}$ obtained above, 
the Ward identity for the QCD virtual corrections to two loops to the process \eqref{eq:process} and  \eqref{eq:Aprocess} with a CP-even 
and CP-odd Higgs boson, respectively,
keeping a non-vanishing Yukawa coupling only for the $b$ quark. Let us emphasize again that no Higgs bremsstrahlung from the $Z$ boson 
is considered here.

\subsection{Derivation of the Ward identity} 
\label{sec:WIderivation}

The classical Lagrangian that encodes all the aforementioned information reads as  
\begin{eqnarray}
 \mathcal{L}_{c} = & - \frac{1}{4} G^a_{\mu\nu}G^{a\mu\nu} + {\overline b}i\gamma^\mu D_\mu b  -J^\mu_{\rm Z} Z_\mu 
 - \lambda_b {\overline b} b H - {\tilde\lambda}_b {\overline b} i \gamma_5 b A \, ,
   \label{eq:lagbhz}
\end{eqnarray}
where $G^a_{\mu\nu}$ denotes the gluon field strength tensor, 
\begin{equation}
 D_\mu = \partial_\mu - i g_s T^a G_\mu^a, \quad J^\mu_{\rm Z} =  g_{V,b} J^\mu  + g_{A,b} J_{5}^\mu\, , \quad
 J^\mu = {\overline b} \gamma^\mu b \, , \,\, J_{5}^\mu = {\overline b} \gamma^\mu \gamma_5 b \, , 
  \label{eq:bnccur}
\end{equation}
and $H, A$ denotes a CP-even and CP-odd Higgs boson, respectively. The kinetic terms of the Higgs bosons and of the $Z$ boson are not listed in  \eqref{eq:lagbhz}.

Performing a  continuous global $U(1)$ transformation
 and applying the Noether theorem to the classical Lagrangian \eqref{eq:lagbhz}, 
 \begin{equation*}
  b(x) \rightarrow e^{i\alpha} b(x) \qquad \text{and} \qquad {\overline b}(x) \rightarrow e^{-i\alpha} {\overline b}(x) \,,
 \end{equation*}
leads to the well known conservation law for the vector current  
\begin{equation}
 \partial^\mu  J_\mu = 0 \, ,
 \label{eq:consvec}
\end{equation}
which holds exactly also at the quantum level.
 Performing the continuous global $U_A(1)$ transformation
\begin{equation}
  b(x) \rightarrow e^{i\alpha \gamma_5} b(x) \qquad \text{and} \qquad {\overline b}(x) \rightarrow {\overline b}(x) e^{i\alpha \gamma_5}  
  \label{eq:axtra}
 \end{equation}
gives rise, at the classical level, to the following equation: 
 \begin{equation}
 \partial^\mu  J_{5\mu} = - 2 \lambda_b {\overline b} i \gamma_5 b H + 2 {\tilde\lambda}_b {\overline b} b A \,.
   \label{eq:consax}
 \end{equation}
If one had kept the $b$ quark massive, then there would be also a $b$-quark mass-dependent term appearing on the right-hand side of \eqref{eq:consax}.
At the quantum level the current $J_{5}^{\mu}$ suffers from the  ABJ anomaly~\cite{Adler:1969gk,Bell:1969ts}, i.e.,~the term $-a_s 
\epsilon^{\mu\nu\rho\sigma}G^a_{\mu\nu}G^a_{\rho\sigma}/2$ appears\footnote{We use the convention $\varepsilon^{0123}=-\varepsilon_{0123}=+1$.}
in addition on  the right-hand side of \eqref{eq:consax}.
Proper renormalization of fields and interactions involved are understood wherever needed.
We will come back to this later.
~\\

Next we derive a relation between the S-matrix element of the process \eqref{eq:process}, and likewise for \eqref{eq:Aprocess}, 
and a respective S-matrix element where the $Z$ boson has been removed. We call this relation a Ward identity, although the axial vector current involved
in this relation is not even partially conserved (cf. eq. \eqref{eq:consax}).

{\it CP-even Higgs boson:} The corresponding S-matrix element is 
\begin{equation}
  \langle  Z(q_1) H(q_2)\,|\, b(p_1) {\overline b}(p_2) \rangle =  (2\pi)^4\delta^{(4)}(q_1+q_2-p_1-p_2){\cal M}_\mu \epsilon_Z^{*\mu} \, ,
  \label{eq:S-TmatelH}
\end{equation}
where $\epsilon_Z^{\mu}$ denotes the polarization vector of the $Z$ boson.
We consider here only the so-called non-anomalous QCD contributions to the S-matrix element \eqref{eq:S-TmatelH}, 
namely eq.~\eqref{eq:consax} will be used without the ABJ anomaly term.

Applying the LSZ reduction formalism~\cite{Lehmann:1954rq} to the S-matrix element \eqref{eq:S-TmatelH} 
one obtains\footnote{With the sign convention for the neutral current interaction chosen in \eqref{eq:lagbhz} the free field equation 
for the $Z$-boson field is $\Big( g^{\mu\nu}(\partial^2 + m_Z^2) - \partial^\mu\partial^\nu \Big) Z_\nu = J^\mu_{\rm Z} \, .$}
 \begin{equation}
 \langle  Z(q_1) H(q_2)| b(p_1) {\overline b}(p_2) \rangle = {\rm disc.} - i \int d^4x\, e^{iq_1\cdot x}\langle H(q_2)|J^\mu_{\rm Z}(x) | b(p_1) {\overline b}(p_2) \rangle \, 
 \epsilon^*_{Z,\mu} \, , 
 \label{eq:LSZ1}
\end{equation}
where disc. denotes disconnected terms that do not contribute here.
Let us further denote
\begin{equation}
 \mathrm{M}^\mu \equiv -i \int d^4x\, e^{iq_1\cdot x}\langle H(q_2)|J^\mu_{\rm Z}(x) | b(p_1) {\overline b}(p_2) \rangle \, .
 \label{eq:deffmu}
\end{equation}
Contracting $\mathrm{M}^\mu$ with the four-momentum of the $Z$ boson we obtain
\begin{eqnarray}
 q_{1,\mu} \, \mathrm{M}^\mu 
  & = &  \int d^4x\,  e^{iq_1\cdot x} \langle H(q_2)|\partial_\mu J^\mu_{\rm Z}(x) | b(p_1) {\overline b}(p_2) \rangle \nonumber \\
 & = & -2 g_{A,b} \lambda_b  (2\pi)^4\delta^{(4)}(q_1+q_2-p_1-p_2) \langle H(q_2)| {\overline b}(0) i \gamma_5 b(0) H(0)| b(p_1) {\overline b}(p_2) \rangle \, , \nonumber \\
 \label{eq:pi-noeth}
\end{eqnarray}
where we used the conservation of the vector current, the divergence of the axial vector current, 
eq.~\eqref{eq:consax} (in the absence of ABJ anomaly), 
and translation invariance.
From eqs.~\eqref{eq:S-TmatelH} - \eqref{eq:pi-noeth} we obtain the following relation which we call a Ward identity:
 \begin{equation}
  q_1^\mu {\cal M}_\mu  = -2 g_{A,b} \lambda_b   \langle H(q_2)| {\overline b}(0) i \gamma_5 b(0) H(0)| b(p_1) {\overline b}(p_2) \rangle \, .
  \label{eq:WardIH}
 \end{equation}
This relation holds to all orders in the QCD coupling (in the absence of the anomalous diagrams) 
with the dynamics specified by the Lagrangian \eqref{eq:lagbhz} in its renormalized form. 
Notice that the kinematics of the matrix element on the right-hand side of this equation 
obeys $p_1+p_2 -q_2=q_1$, where $q_1^2 = m_Z^2$. 
This is due to the external momentum insertion $q^{\mu}_1$ introduced 
by the local composite operator ${\overline b}(0) i \gamma_5 b(0) H(0)$ (which is understood to be normal-ordered).
To lowest order in perturbation theory one gets
 \begin{equation}
  \langle H(q_2)| {\overline b}(0) i \gamma_5 b(0) H(0)| b(p_1) {\overline b}(p_2) \rangle = {\overline v}(p_2) i \gamma_5 u(p_1) \, .
  \label{eq:HverLO}
 \end{equation}
We remark that the overall sign of the right-hand side follows from the definition of the initial two-fermion state. 
The three-point pseudoscalar vertex \eqref{eq:HverLO} represents an incoming $b$ and ${\overline b}$ quark with four-momenta $p_1$ and $p_2$, 
respectively, and an outgoing Higgs boson with four-momentum $q_2$. It is depicted in figure~\ref{dia:ward_rhs}. ~\\

{\it CP-odd Higgs boson:}  The corresponding S-matrix element is 
 \begin{equation}
  \langle  Z(q_1) A(q_2)\,|\, b(p_1) {\overline b}(p_2) \rangle =  (2\pi)^4\delta^{(4)}(q_1+q_2-p_1-p_2){\cal M}^A_\mu \epsilon_Z^{*\mu} \, .
  \label{eq:S-TmatelA}
 \end{equation} 
  The respective Ward identity for \eqref{eq:S-TmatelA} can be derived in completely analogous fashion using  \eqref{eq:consax}.
  We obtain 
  \begin{equation}
   q_1^\mu {\cal M}^A_\mu  = 2 g_{A,b} {\tilde\lambda}_b   \langle A(q_2)| {\overline b}(0) b(0) A(0)| b(p_1) {\overline b}(p_2) \rangle \, .
  \label{eq:WardIA}
  \end{equation}
 Evaluating the right-hand side of \eqref{eq:WardIA} to lowest order we get 
\begin{equation}
  \langle A(q_2)| {\overline b}(0) b(0) H(0)| b(p_1) {\overline b}(p_2) \rangle = {\overline v}(p_2) u(p_1) \, .
  \label{eq:AverLO}
 \end{equation}
The remarks made below \eqref{eq:WardIH} and \eqref{eq:HverLO} apply also here.

\subsection{Checking the Ward identity} 
\label{sec:WIcheck}

Next we verify the Ward identity \eqref{eq:WardIH} for a CP-even Higgs boson $H$
at the level of the UV renormalized and IR subtracted finite remainders in four dimensions. 
From the discussion of section~\ref{suse:intva}
 we see that the virtual (non-anomalous) two-loop QCD corrections to the amplitude of the process \eqref{eq:process} are given by
\begin{align} 
\label{eq:bbZHamplitude_FFR}
\mathcal{M}_{vec}^{\mu} &= 
    \left(F_{1,HZ} + F_{1,ZH} \right) \, \bar{v}(p_2) \, u(p_1) \, p_1^{\mu} 
    \,+\, \left(F_{2,HZ} + F_{2,ZH} \right) \, \bar{v}(p_2) \, u(p_1) \, p_2^{\mu} \nonumber\\
    &\,+\, \left(F_{3,HZ} + F_{3,ZH} \right) \, \bar{v}(p_2) \, u(p_1) \, q_1^{\mu}
    \,+\, \left(F_{4,HZ} + F_{4,ZH} \right) \, \bar{v}(p_2) \,\gamma^{\mu} \slashed{q}_1 u(p_1) \,, \\
\mathcal{M}_{axi(ns)}^{\mu} &= 
    \left(F_{1,HZ} - F_{1,ZH} \right) \, \bar{v}(p_2) \,\gamma_5\, u(p_1) \, p_1^{\mu} 
    \,+\, \left(F_{2,HZ} - F_{2,ZH} \right) \, \bar{v}(p_2) \,\gamma_5\, u(p_1) \, p_2^{\mu} \nonumber\\
    &\,+\, \left(F_{3,HZ} - F_{3,ZH} \right) \, \bar{v}(p_2) \,\gamma_5\, u(p_1) \, q_1^{\mu}
    \,+\, \left(F_{4,HZ} - F_{4,ZH} \right) \, \bar{v}(p_2) \,\gamma^{\mu} \slashed{q}_1 \gamma_5 u(p_1) \,, 
\end{align}
where the previously calculated analytic expressions of  the vector form factors $F_{i,ZH}$ and $F_{i,HZ}$ 
will be inserted and all spinor products are to be evaluated in four dimensions.

\begin{figure}[htbp]
\begin{center}
\includegraphics[scale=0.50]{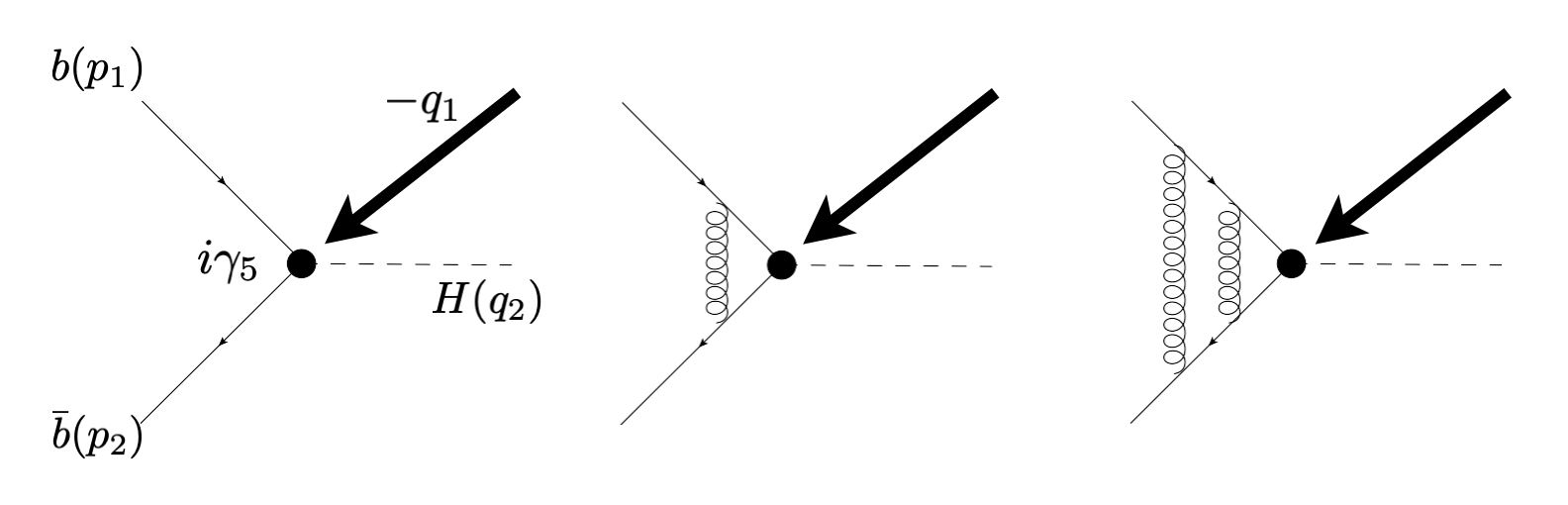}
\caption{Example diagrams at tree level, one-loop and two-loop corresponding to the right-hand side of the Ward identity \eqref{eq:WardIH}. 
The solid arrow indicates the insertion of the external momentum $-q_1$  into the vertex.}
\label{dia:ward_rhs}
\end{center}
\end{figure}

The left-hand side of the Ward identity \eqref{eq:WardIH} is obtained by contracting the  amplitudes \eqref{eq:bbZHamplitude_FFR} 
with the four-momentum of the $Z$ boson.
 We get
\begin{align} 
q_1 \cdot \mathcal{M}_{vec} &= \bar{v}(p_2) \, u(p_1) \, 
    \Bigg(\left(F_{1,HZ} + F_{1,ZH} \right) \frac{m_Z^2-t}{2} 
    \,+\, \left(F_{2,HZ} + F_{2,ZH} \right) \frac{m_Z^2-u}{2} \nonumber   \\
    &\,+\, \left(F_{3,HZ} + F_{3,ZH} \right) \, m_Z^2
    \,+\, \left(F_{4,HZ} + F_{4,ZH} \right) \, m_Z^2 \Bigg), \label{eq:WI_LHSV}\\
q_1 \cdot \mathcal{M}_{axi(ns)} &= \bar{v}(p_2) \,\gamma_5\, u(p_1) \, 
    \Bigg(\left(F_{1,HZ} - F_{1,ZH} \right) \frac{m_Z^2-t}{2} 
    \,+\, \left(F_{2,HZ} - F_{2,ZH} \right) \frac{m_Z^2-u}{2} \nonumber\\
    &\,+\, \left(F_{3,HZ} - F_{3,ZH} \right) \, m_Z^2
    \,+\, \left(F_{4,HZ} - F_{4,ZH} \right) \, m_Z^2 \Bigg) \, . \label{eq:WI_LHSA}
\end{align}
The form factors $F_{i,ZH}$  and $F_{i,HZ}$  can be UV-renormalized and IR-subtracted as outlined
in detail in ref.~\cite{Ahmed:2019udm}, which subsequently leads to their respective finite remainders in four dimensions. 
Inserting these finite remainders of $F_{i,ZH}$ and $F_{i,HZ}$ into \eqref{eq:WI_LHSV} and \eqref{eq:WI_LHSA}, respectively,  we 
obtain the finite remainders of the left-hand side \eqref{eq:WardIH} in four dimensions. 
In particular we have verified that $q_1 \cdot \mathcal{M}_{vec} = 0$. (This holds, of course, already before renormalization and subtraction.)

The right-hand side of \eqref{eq:WardIH} consists, up to the factor $-2 g_{A,b} \lambda_b$, of the tree-level three-point pseudoscalar vertex \eqref{eq:HverLO}
and its one-loop and two-loop QCD virtual corrections, albeit with a special kinematic configuration as explained above. 
Example diagrams are shown in figure~\ref{dia:ward_rhs}. 
All contributions are proportional to the Lorentz structure $\bar{v}(p_2) \,\gamma_5\, u(p_1)$. 
We used here an anticommuting $\gamma_5^{AC}$ for the pseudoscalar vertex. 
After carrying out the UV renormalization and IR subtraction of these QCD virtual corrections, 
we obtained an expression that agrees analytically with the aforementioned finite remainder of the left-hand side of \eqref{eq:WI_LHSA}.

In the case of a CP-odd Higgs boson, where the vector and axial form factors, i.e.~$F^A_{i,vec}$ and $F^A_{i,axi(ns)}$, are given by \eqref{eq:fofacA} in terms of $F_{i,ZH}$ and $F_{i,HZ}$, it is straightforward to verify the Ward identity \eqref{eq:WardIA} 
in completely analogous fashion.
To be more specific, with the parameterization of the Yukawa couplings as in \eqref{eq:lagbhz}, 
we have $F^A_{j,vec} = i\,  F_{j,vec}$ and $F^A_{j,axi(ns)} = i\, F_{j,axi(ns)}$. 
The two sets of form factor-factor decomposition bases of the amplitudes involving $H$ and $A$ differ just by an additional $\gamma_5$ sandwiched between spinors.
Likewise, the r.h.s. of the Ward identities \eqref{eq:WardIH} and \eqref{eq:WardIA}
differ by the factor $i \gamma_5$ sandwiched between spinors, 
as far as the Lorentz structure is concerned. 
Thus the form factors \eqref{eq:fofacA} fulfill the Ward identity \eqref{eq:WardIA}.


\section{Top-Yukawa coupling dependent top-loop contributions to $q \bar{q} \rightarrow ZH$ in Higgs effective theory}
\label{sec:qqZHinHEFT}

\begin{figure}[htbp]
\begin{center}
\includegraphics[scale=0.50]{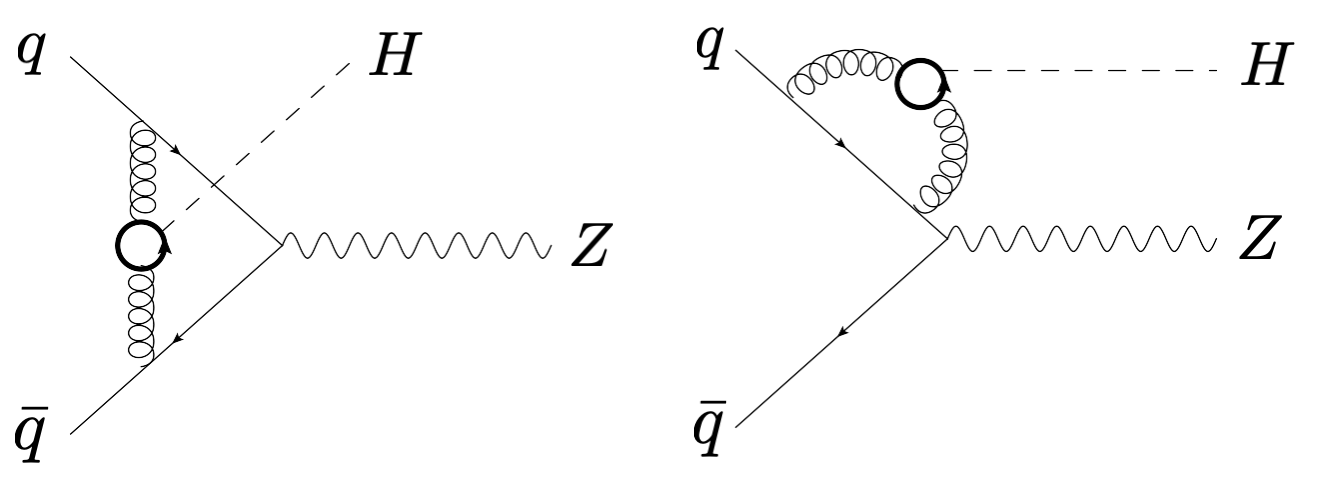}
\caption{The two-loop class-I diagrams of the top contributions to $q\bar{q} \rightarrow ZH$ proportional to
 $\lambda_t$. The thick solid lines denote the massive top quark.}
\label{dia:LO_full}
\end{center}
\end{figure}

Starting from $\mathcal{O}(\alpha_s^2)$ in QCD, a new class of two-loop diagrams contributes to the quark antiquark annihilation 
initiated $Z h$ process where the Higgs boson couples directly to a closed top-quark loop, and hence 
the top-quark Yukawa coupling  gets involved. 
This was studied for $h=H$ for instance in ref.~\cite{Brein:2003wg,Brein:2011vx} by making use of asymptotic expansions in the heavy-top limit. 
In ref.~\cite{Brein:2011vx}, the two-loop virtual top-quark contributions to $q\bar{q} \rightarrow ZH$ proportional to  $\lambda_t$ were classified 
into two sets: class-I (with examples shown in figure~\ref{dia:LO_full}) 
and class-II, depending on whether the $Z$ boson couples to the external light quark in the initial state or to  the virtual 
top-quark loop (from which the Higgs boson is radiated), giving rise to different electroweak
coupling factors. 

These contributions were not covered in ref.~\cite{Ahmed:2019udm} where the 
computations were made in $n_f = 5$ massless QCD with a non-vanishing Yukawa coupling for the $b$ quark only
(as discussed in the preceding sections except for the section~\ref{suse:anomalousdiagrams} where the top-induced triangle diagrams are included in addition). 
As the second part of the work presented in this article, 
 we compute the contributions of the class-I diagrams to $\mathcal{O}(\alpha_s^3)$ in 
the heavy-top limit using the Higgs effective field theory (HEFT) to two-loop order. 
 We confine ourselves in the following to a scalar Higgs boson $h=H$. 
Our motivation for presenting these contributions here in some detail is a problem in applying a non-anticommuting $\gamma_5$ 
that we encountered in the computation of this class of contributions.

We parameterize in the following the general Yukawa coupling $\lambda_t$ of a CP-even Higgs boson to the top quark by
 \begin{equation}
  \label{eq:gen:tYuk}
  \lambda_t = - c_t \frac{m_t}{v} \, ,
 \end{equation}
where $-m_t/v$ with $v=246$ GeV is the SM top-Yukawa coupling and the dimensionless parameter $c_t$ depends on the
 specific Higgs model.

\subsection{HEFT and UV renormalization}
\label{sec:qqZHinHEFT_uv}

\begin{figure}[htbp]
\begin{center}
\includegraphics[scale=0.55]{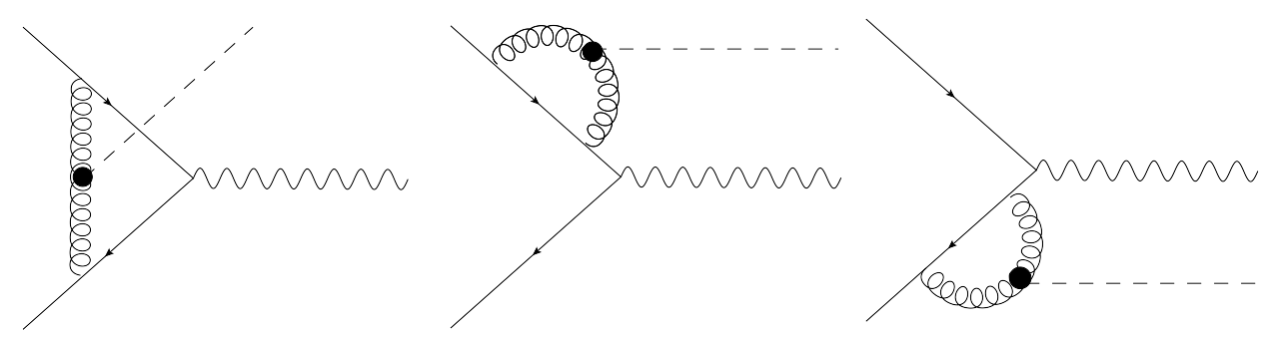}
\caption{The class-I diagrams of the top-quark
 contributions to $q\bar{q} \rightarrow ZH$ in the limit $m_t \rightarrow \infty$ in the leading power approximation. 
The black blob indicates the effective $Hgg$ vertex. }
\label{dia:LO_HEFT}
\end{center}
\end{figure}

In ref.~\cite{Brein:2011vx} it was pointed out that applying the heavy-mass expansion procedure 
to the class-I diagrams of figure \ref{dia:LO_full} which leads to terms similar to
those depicted in figure~\ref{dia:mtinfinity1}, albeit with $Z$ and $H$ exchanged, 
the expanded terms featuring the effective $q\bar{q}ZH$ or $q\bar{q}H$ vertex vanish to leading power in $1/m_t$. 
Only the terms that involve the effective Higgs-gluon-gluon ($Hgg$) vertex contribute in the infinite $m_t$ limit.
Thus, the leading approximation in powers of $1/m_t$ of the diagrams in figure~\ref{dia:LO_full} can
be described with the Higgs effective Lagrangian (see e.g.~ref.\cite{Kniehl:1995tn}) 
where the top quark is integrated out. 
The diagrams of figure~\ref{dia:LO_full} are then reduced to the one-loop diagrams 
shown in figure~\ref{dia:LO_HEFT}.

However, as will become clear at the end of the next section, it turns out that the validity 
of this point depends on the particular $\gamma_5$ prescription in use. 
To be more specific, it is not true for the axial current regularized using a non-anticommuting $\gamma_5$,
which can be deduced from our computations described below.
This is one of the key results conveyed through the remaining sections.

In the Higgs effective theory, where the top-quark degrees of freedom are integrated out, 
the Lagrangian density that encapsulates the interaction between the scalar Higgs boson and gluons 
is given by (neglecting terms that are not relevant here)
\begin{align}
\label{eq:HEFT-Lag}
    {\cal L}_{\rm heff}= -\frac{1}{4} c_t~C_{H}  \frac{H}{v} G^a_{\mu\nu} G^{a,\mu\nu}\,,
\end{align}
where $c_t$ and $v$ are defined in and below eq.~\eqref{eq:gen:tYuk}, respectively,
 and $C_{H}$  denotes the Wilson coefficient 
 that is determined for a Standard Model Higgs boson by matching the effective $n_f=5$ flavor theory to the full $(n_f+1)$-flavor 
 theory order-by-order in the QCD coupling. 
To second order in $a_s(\mu_R^2)$ it is given by \cite{Kramer:1996iq,Chetyrkin:1997iv}
\begin{align}
\label{eq:Wilson-res}
    C_{H}\left( a_s(\mu_R^2)\right) =& 
     -\frac{4a_s}{3} \bigg(1 + a_s \Big\{5C_A-3C_F\Big\} \bigg)\,,
\end{align}
where $\mu_R$ is the renormalization scale and $C_A,~C_F$ denote the quadratic Casimir operators of the SU($N_c$) color gauge group 
in the adjoint and fundamental representations, respectively.

The effective operator \eqref{eq:HEFT-Lag} must be renormalized, 
in addition to performing the QCD coupling renormalization (done in the ${\overline{\rm MS}}$ scheme), in order 
to get rid of all the UV poles appearing in the scattering amplitudes. 
This is achieved by
\begin{align}
\label{eq:Operator-renorm}
    &\big[{\cal L}_{\rm heff}\big]_R=Z_{H} {\cal L}_{\rm heff}
\end{align}
with the operator renormalization constant~\cite{Nielsen:1975ph,Spiridonov:1988md,Kataev:1981gr}
\begin{align}
\label{eq:opZH}
    Z_{H}=1-a_s \bigg( \frac{1}{\epsilon}\beta_0\bigg)+a_s^2 \bigg(\frac{\beta_0^2}{\epsilon^2}-\frac{\beta_1}{\epsilon}\bigg)\,.
\end{align}
The coefficients of the QCD $\beta$-function are given by
\begin{align}
\label{eq:beta0}
    \beta_0&=\frac{11}{3}C_A-\frac{2}{3}n_f\,,\nonumber\\
    \beta_1&=\frac{34}{3} C_A^2-2 C_F n_f -\frac{10}{3} C_A n_f\,.
\end{align}

\begin{figure}[htbp]
\begin{center}
\includegraphics[scale=0.47]{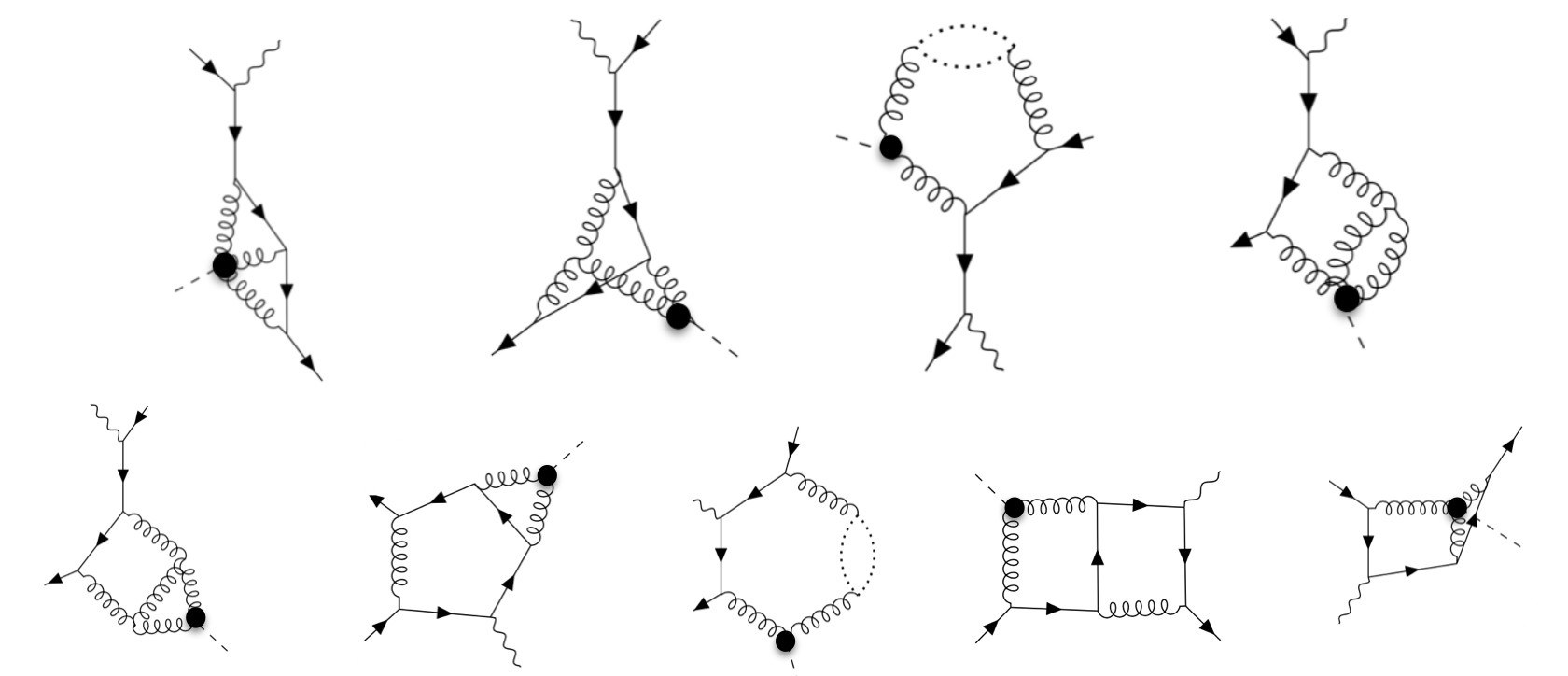}
\caption{Examples of the class-I two-loop diagrams proportional 
to $\lambda_t$ involving only an open massless quark line $q$ for $q\bar{q} \rightarrow ZH$ in HEFT.}
\label{dia:nonsinglet_HEFT}
\end{center}
\end{figure}

\subsection{Form factors of the class-I contributions using an anticommuting $\gamma_5$}
\label{sec:qqZHinHEFT_vFF}

\begin{figure}[htbp]
\begin{center}
\includegraphics[scale=0.47]{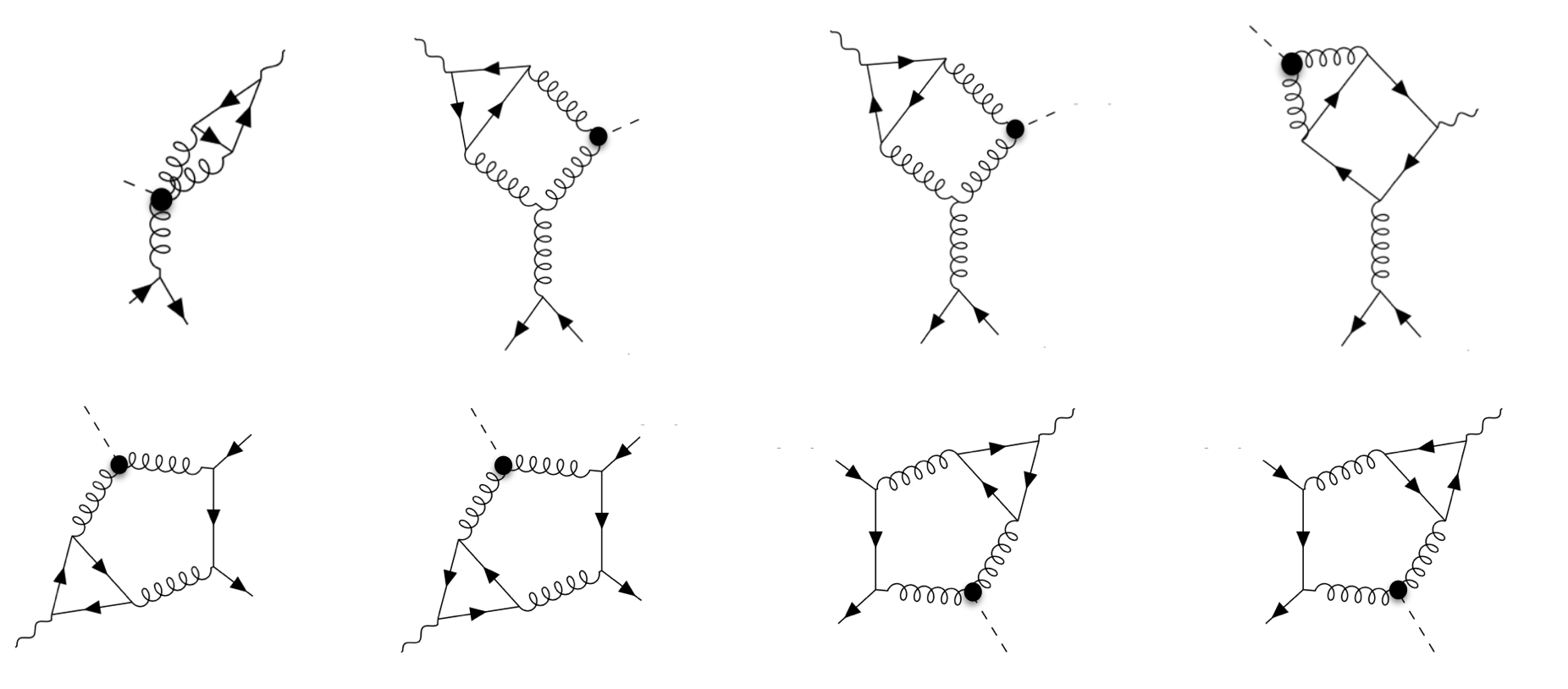}
\caption{The two-loop diagrams proportional to $\lambda_t$ for $q\bar{q} \rightarrow ZH$ with an open massless quark line $q$ and a quark loop in HEFT.
The first four diagrams vanish because of color conservation.}
\label{dia:singlet_HEFT}
\end{center}
\end{figure}

We consider now the class-I $\lambda_t$-dependent contributions to $q{\bar q}\to Z H$ in HEFT. 
The leading-order contributions are depicted in figure~\ref{dia:LO_HEFT}. 
Examples of the two-loop non-anomalous QCD corrections that involve only an open massless quark line $q$ which the $Z$ boson couples to are shown in figure~\ref{dia:nonsinglet_HEFT}, 
and the two-loop QCD corrections that involve in addition a closed quark loop which the $Z$ boson couples to are displayed in figure~\ref{dia:singlet_HEFT}.
We denote the corresponding contributions to the amplitude by ${\cal A}$  in order to distinguish it from the contributions discussed in section~\ref{sec:AfromV}:

\begin{align} 
\label{eq:qqZHamplitude}
    \mathcal{A} &= c_t~ \bar{v}(p_2) \, \mathbf{\Gamma}^{\mu}_E \, u(p_1) \, \varepsilon^{*}_{\mu}(q_1) \nonumber\\
    &= c_t~g_{V,q}~\bar{v}(p_2) \, \mathbf{\Gamma}^{\mu}_{vec,E} \, u(p_1) \, \varepsilon^{*}_{\mu}(q_1) 
    \,+\, c_t~\bar{v}(p_2) \, \mathbf{\Gamma}^{\mu}_{axi,E} \, u(p_1) \, \varepsilon^{*}_{\mu}(q_1)\nonumber\\
    &\equiv c_t~g_{V,q}~{\cal A}_{vec}+ c_t~{\cal A}_{axi}\,,
\end{align}
where the suffix $E$ indicates the use of the Higgs effective Lagrangian \eqref{eq:HEFT-Lag},  and we have factored out $c_t$ defined in \eqref{eq:gen:tYuk} and the vector coupling $g_{V,q}$ 
of the light quark $q$ to the $Z$ boson. As to the dependence on axial vector couplings see eq.~\eqref{eq:Aaxisns} below.

The vector part of the amplitude can be decomposed in terms of linearly independent 
and complete Lorentz structures in D dimensions as
\begin{align} 
\label{eq:FFdecomp-qqZH-vec}
\bar{v}(p_2) \, \mathbf{\Gamma}^{\mu}_{vec,E} \, u(p_1) &= 
    {\cal F}_{1,vec} \, \bar{v}(p_2) \,\slashed{q_1}\, u(p_1) \, p_1^{\mu} 
    \,+\, {\cal F}_{2,vec} \, \bar{v}(p_2) \,\slashed{q_1}\, u(p_1) \, p_2^{\mu} \nonumber\\
    &\,+\, {\cal F}_{3,vec} \, \bar{v}(p_2) \,\slashed{q_1}\, u(p_1) \, q_1^{\mu}
    \,+\, {\cal F}_{4,vec} \, \bar{v}(p_2) \,\gamma^{\mu} u(p_1) \,.
\end{align}
The decomposition reflects the chiral conservation along the massless quark line.
The ${\cal F}_{i,vec}$ are the
vector form factors which are computed by applying the corresponding projectors 
to the respective Feynman diagrams. These vector form factor projectors read in D dimensions:
\begin{align} 
\label{eq:FFD_vec_projectors}
    {\mathbb{P}}^{\mu}_{1} &= \bar u(p_1)
    \Big\{
        (2 - D) (m_z^2 - u)^2 p_1^{\mu} 
        + \Big(m_z^4 (D-4) + (D-4) t u - m_z^2 (2 s (D-3) \nonumber\\
        &~~~~~~~~~~~~+ (D-4) (t + u))\Big) p_2^{\mu} + s (D-2)  (m_z^2 - u) q_1^{\mu} \nonumber\\
        &~~~~~~~~~~~~+ (m_z^2 - u) (m_z^4 + t u - m_z^2 (s + t + u)) \gamma^{\mu} 
    \Big\}
        v(p_2) \frac{1}{{\cal K}^2}\,,\nonumber\\
    {\mathbb{P}}^{\mu}_{2} &= \bar u(p_1)
    \Big\{
        \Big(m_z^4 (D-4) + (D-4) t u - m_z^2 (2 (D-3) s + (D-4) (t + u))\Big) p_1^{\mu}\nonumber\\
        &~~~~~~~~~~~~+ (2 - D) (m_z^2 - t)^2 p_2^{\mu}
        + (D-2) s (m_z^2 - t) q_1^{\mu} \nonumber\\
        &~~~~~~~~~~~~+ \Big( (m_z^2 - t) (m_z^4 + t u - m_z^2 (s + t + u))\Big) \gamma^{\mu}
    \Big\}
        v(p_2) \frac{1}{{\cal K}^2}\,,\nonumber\\
    {\mathbb{P}}^{\mu}_{3} &= \bar u(p_1)
    \Big\{
        (D-2) (m_z^2 - u) p_1^{\mu}
        + (D-2) (m_z^2 - t) p_2^{\mu} 
        +(2 - D) s q_1^{\mu} \nonumber\\
        &~~~~~~~~~~~~+ \Big( -m_z^4 - t u + m_z^2 (s + t + u) \Big) \gamma^{\mu} 
    \Big\}
    v(p_2)\frac{1}{{\cal K}^2}\,,\nonumber\\
    {\mathbb{P}}^{\mu}_{4} &= \bar u(p_1)
    \Big\{
         (m_z^2 - u)p_1^{\mu}
        + (m_z^2 - t) p_2^{\mu} 
        - s q_1^{\mu} \nonumber\\
        &~~~~~~~~~~~~+ \Big(-m_z^4 - t u + m_z^2 (s + t + u) \Big) \gamma^{\mu} 
    \Big\}
    v(p_2)\frac{1}{{\cal K}} \, ,
\end{align}
where
\begin{align}
    {\cal K} = 2 s (D-3) (m_z^4 +t u - m_z^2 (s + t+ u))\,.
\end{align}
We note that, by construction, the index contraction between these projectors \eqref{eq:FFD_vec_projectors} 
and the $\varepsilon^{*}_{\mu}$-stripped amplitude is to be done with the D-dimensional 
 space-time metric tensor $g_{\mu \nu}$.

The UV renormalization of the amplitude \eqref{eq:FFdecomp-qqZH-vec} proceeds 
as explained in section~\ref{sec:qqZHinHEFT_uv}. 
The technical aspects of the computation of these vector form factors closely follow 
the steps as explained in detail in ref.~\cite{Ahmed:2019udm}.
We remark that the two-loop amplitudes (or form factors) \eqref{eq:FFdecomp-qqZH-vec} 
involves 117 master integrals,
for which we take the analytic expressions computed in ref.~\cite{Gehrmann:2015ora}  
available in \hepforge~\cite{hepforge} in computer readable format. 
The UV renormalized vector form factors ${\cal F}_{i,vec}$ at one as well as two loops were checked for exhibiting 
the universal infrared structures~\cite{Catani:1998bh,Sterman:2002qn,Becher:2009cu,Becher:2009qa,Gardi:2009qi}. 
This serves as a strong check of our computations.

The expansion of the UV renormalized vector form factors in powers of $a_s$ is defined formally by
\begin{align}
\label{eq:calF-expand}
    {\cal F}_{i,vec}= \sum_{l=1}^{\infty} a^l_s(\mu_R) {\cal F}_{i,vec}^{(l)}\,.
\end{align}
The analytic results of these UV renormalized vector form factors are too lengthy 
to be presented here, but they can be provided upon demand 
from the authors\footnote{In particular, the file containing the four UV renormalized 
two-loop amplitudes ${\cal M}^{[j],(2)}$ is about 12 MB.}.

 We decompose the  amplitude ${\cal A}_{axi}$ associated with the axial vector current as follows:
 \begin{equation}
  \label{eq:Aaxisns}
  {\cal A}_{axi} = g_{A,q}~{\cal A}_{axi(ns)} + g_{A,b}~{\cal A}_{axi(s)} \, ,
 \end{equation}
 where the  contribution ${\cal A}_{axi(ns)}$ covers the one- and two-loop diagrams of
 figures~\ref{dia:LO_HEFT} and~\ref{dia:nonsinglet_HEFT},
 respectively, while the term ${\cal A}_{axi(s)}$ results from the non-vanishing two-loop diagrams of figure~\ref{dia:singlet_HEFT}. 
The dependence of this term on $g_{A,b}$ only will be discussed in section~\ref{suse:UVFSiC} below.

 The axial components of the form factors are defined by decomposing ${\cal A}_{axi}$ 
 in analogy to \eqref{eq:FFdecomp-qqZH-vec} as follows 
(we drop here the additional subscript `ns' or `s' for ease of notation):
 \begin{align} 
\label{eq:FFD_axi}
{\mathcal{A}}^{\mu}_{axi} &\equiv  
    \mathcal{F}_{1,axi} \, \bar{v}(p_2) \,\slashed{q_1} \gamma_5\, u(p_1) \, p_1^{\mu} 
    \,+\, \mathcal{F}_{2,axi} \, \bar{v}(p_2) \,\slashed{q_1} \gamma_5\, u(p_1) \, p_2^{\mu} \nonumber\\
    &\,+\, \mathcal{F}_{3,axi} \, \bar{v}(p_2) \,\slashed{q_1} \gamma_5\, u(p_1) \, q_1^{\mu}
    \,+\, \mathcal{F}_{4,axi} \, \bar{v}(p_2) \,\gamma^{\mu} \gamma_5 \, u(p_1) \,,
\end{align}
and these form factors are expanded in powers of $a_s$ in analogy to the vector counterparts
in \eqref{eq:calF-expand}.

For the computation of the non-anomalous part ${\cal A}_{axi(ns)}$, respectively the form factors
$\mathcal{F}_{2,axi(ns)}$, 
we use an anticommuting $\gamma_5^{AC}$ in D dimensions. 
Then  ${\cal A}_{axi(ns)}$ respects chiral invariance  which implies
that 
\begin{align}
\label{eq:Ax-equal-Vec-Eff}
    {\cal F}_{i,axi(ns)}^{(l)}={\cal F}_{i,vec}^{(l)}\,, \quad i=1,2,3,4.
\end{align}
In order to check eq.~\eqref{eq:Ax-equal-Vec-Eff} we derive the projectors that correspond
to the decomposition \eqref{eq:FFD_axi} (with an anticommuting $\gamma_5^{AC}$).
With these projectors we computed the ${\cal F}^{(l)}_{i,axi(ns)}$ at one and two loops ($l=1,2$)
and confirm  eq.~\eqref{eq:Ax-equal-Vec-Eff}.

For the anomalous two-loop diagrams, where only axial vector part survives, 
we employ a non-anticommuting $\gamma_5$ in dimensional regularization~\cite{tHooft:1972tcz,Breitenlohner:1977hr,Akyeampong:1973xi,Larin:1991tj,Larin:1993tq}.
 Our results will be discussed in section~\ref{suse:UVFSiC}. 
 First we cross-check whether the correct non-anomalous axial form factors (i.e., those given by eq.~\eqref{eq:Ax-equal-Vec-Eff})
are obtained using a non-anticommuting $\gamma_5$ as employed in the literature.
Here we found something quite intriguing, which we now turn to in the following section.


\section{A pitfall in applying a non-anticommuting $\gamma_5$ to $\lambda_t$-dependent 
contributions to $q\bar{q} \rightarrow ZH$ in HEFT}
\label{sec:G5pitfall}

In our attempt to compute the two-loop QCD contributions to the 
class-I diagrams\footnote{A comment about the class-II contributions will be given at the end of the following subsection.} in HEFT, 
shown in figure~\ref{dia:LO_HEFT}, we noticed a technical pitfall in applying a non-anticommuting $\gamma_5$.
To our surprise, this $\gamma_5$ issue  manifests itself already in the leading-order (LO) contributions, i.e.~the one-loop diagrams of figure~\ref{dia:LO_HEFT}. 

\subsection{The class-I axial form factors computed using  a non-anticommuting $\gamma_5$}
\label{sec:result_asitis}

Our computation of the axial vector form factors, ${\cal F}_{i,axi}$, as 
introduced in \eqref{eq:FFD_axi}, using a non-anticom\-muting $\gamma_5$ follows closely 
ref.~\cite{Ahmed:2019udm}, especially with regard to the axial form factor decomposition (although a new set of chirality-preserving form factor decomposition basis is needed here). 
The non-anticommuting $\gamma_5$ in dimensional regularization is defined by~\cite{tHooft:1972tcz,Breitenlohner:1977hr}
\begin{align}
\label{eq:gamma5}
	\gamma_5=-\frac{i}{4!}\varepsilon_{\mu\nu\rho\sigma}\gamma^{\mu}\gamma^{\nu}\gamma^{\rho}\gamma^{\sigma}\,,
\end{align}
with the treatment of the Levi-Civita symbol $\varepsilon_{\mu\nu\rho\sigma}$ (as well as the axial vector current) done according to refs.\cite{Akyeampong:1973xi,Larin:1991tj,Zijlstra:1992kj}.
The usage of this definition has profound implications in high-order computations involving an axial vector current in dimensional regularization. 
One of the main messages conveyed through refs.~\cite{Chen:2019wyb,Ahmed:2019udm} is that even if 
the loop amplitudes are not defined or regularized strictly in the \textquotesingle t Hooft-Veltman scheme, expressions for projectors derived in four dimensions are still sufficient and lead to correct results (for physical observables), irrespective of whether the quantity projected out is a form factor or a linearly polarized amplitude.
This is particularly helpful in evaluating loop amplitudes that involve axial vector currents (if a non-anticommuting $\gamma_5$ is used).

If we use an anticommuting $\gamma_5$, then as just discussed in section \ref{sec:qqZHinHEFT_vFF}, we get
\begin{align} 
\label{eq:FFD_vEa_AC5}
\mathcal{F}_{i,vec} = \mathcal{F}^{\mathrm{AC}}_{i,axi (ns)} 
\end{align}
for $i =  1, 2, 3, 4$ to two loops (leaving the anomalous diagrams aside),
where the superscript AC indicates the use of an anticommuting $\gamma_5$.
Accordingly, conservation of the non-singlet light-quark current implies
\begin{align} 
\label{eq:WI_AC5}
q_{1,\mu} \, \bar{v}(p_2) \, \mathbf{\Gamma}^{\mu}_{vec,E} \, u(p_1) &= 0\,, \nonumber\\
q_{1,\mu} \, \mathcal{A}^{\mu}_{axi (ns)} &= 0\, ,
\end{align}
where $\mathcal{A}^{\mu}_{axi (ns)}$ is given by \eqref{eq:FFD_axi} with  
$\mathcal{F}^{\mathrm{AC}}_{i,axi (ns)}$ inserted.
 Alternatively, if we use the non-anticommuting $\gamma_5$ (NAC) \eqref{eq:gamma5}, 
then the vector part does not change of course, but for the axial FF we find at 
LO in the four-dimensional limit ($\epsilon = (4-D)/2 = 0$):
\begin{align} 
\label{eq:FFD_vEa_NAC5_s}
\mathcal{F}^{\mathrm{NAC}}_{i,axi (ns)} &= \mathcal{F}_{i,vec} \, , \qquad i=1,2,3, \nonumber\\
\end{align}
and
\begin{align} 
\label{eq:FFD_vEa_NAC5_d}
 \mathcal{F}^{\mathrm{NAC}}_{4,axi (ns)} & \neq \mathcal{F}_{4,vec}\, .
\end{align}
Consequently, this difference in the fourth form factor leads to the violation of the 
Ward identity
\begin{align} 
\label{eq:WI_NAC5}
q_{1,\mu} \, \mathcal{A}^{\mu,{\rm NAC}}_{axi (ns)} & \neq  0\, 
\end{align}
already at LO.

\begin{figure}[htbp]
\begin{center}
\includegraphics[scale=0.45]{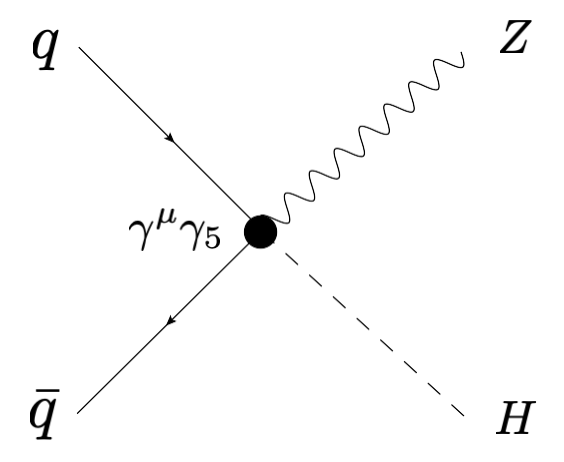}
\caption{The amendment term \eqref{eq:amendO}.}
\label{dia:qqZHoperator}
\end{center}
\end{figure}

With the explicit analytic expressions of these form factors at hand, we find that the expected relations \eqref{eq:Ax-equal-Vec-Eff} 
can be restored at LO if we introduce an additional amendment term, i.e. subtraction term, that is to be added to  $\mathcal{A}^{\mu,{\rm NAC}}_{axi (ns)}$. 
 We denote this term by
\begin{align} 
\label{eq:amendO} 
\mathcal{J}^{\mu,{\rm NAC}}
\equiv 
\mathrm{Z}_5^{h}(a_s)\, \mathrm{\mathbf{C}} \, 
\Big(\bar{v}(p_2) \,\left[\gamma^{\mu} \gamma_5\right]_{L} \, u(p_1) \Big)\, , 
\end{align}
where the constant factor $\mathrm{\mathbf{C}} \equiv a_s \left(-4 C_F\right) \frac{C_H}{v} $ collects the overall perturbative power $a^2_s$ of the LO amplitude (and the effective coupling prefactors) and $\left[\gamma^{\mu} \gamma_5\right]_{L}$ denotes the axial vector current matrix renormalized according to the prescription~\cite{Larin:1991tj,Larin:1993tq}. 
(At this order no renormalization of refs.~\cite{Larin:1991tj,Larin:1993tq} gets involved.)
The additional renormalization constant $\mathrm{Z}_5^{h}(a_s) = 1 + \mathcal{O}(a_s)$ has an 
 expansion in $a_s$ which we will determine explicitly to the first order in $a_s$ from our next-to-leading-order (NLO) computation to be presented later.  
Diagrammatically, this amendment term can be viewed as introducing a four-point \textit{local composite operator} 
corresponding to diagram~\ref{dia:qqZHoperator}
with a multiplicative factor $\mathrm{Z}_5^{h}(a_s)$ to be determined order by order in $a_s$.
With the extra amendment term \eqref{eq:amendO}, the desired properties are restored:
$\mathcal{F}^{\mathrm{NAC}}_{i,axi (ns)} = \mathcal{F}_{i,vec}$ holds for all four form factors and 
$q_{1,\mu} \, \mathcal{A}^{\mu,\mathrm{NAC}}_{axi (ns)} = 0$ is then also fulfilled.~\\

 The source of the inequality \eqref{eq:FFD_vEa_NAC5_d} can be traced at  LO to be solely due to the one-loop box diagram,  i.e., the left-most one in figure~\ref{dia:LO_HEFT}, while the contributions of the two triangle one-loop LO diagrams respect  $\mathcal{F}^{\mathrm{NAC}}_{i,axi (ns)} = \mathcal{F}_{i,vec}$ $(i=1,2,3,4)$ in the four-dimensional limit.
We emphasize that each of the three LO diagrams of figure~\ref{dia:LO_HEFT} is finite.
However, the Feynman amplitude of the finite one-loop box diagram contains terms that are separately divergent.
The expressions obtained using on the one hand an anticommuting $\gamma_5$ and on the other hand, a non-anticommuting one, 
lead to different $D$-dependent coefficients in front of these 
divergent terms, with differences being suppressed by at least one power in (D-4). 
The crucial point is that the (D-4) difference between these two expressions is not an overall prefactor.
It is then not hard to conceive that some non-vanishing evanescent anti-commutators are generated when one shifts the non-anticommuting $\gamma_5$ 
from inside the loop correction of the axial vector vertex to the outside of the loop, which then leads to the observed discrepancy.

\subsection{The NLO QCD corrections}
\label{sec:result_asitisNLO}

\subsection*{UV renormalization of the non-anomalous HEFT diagrams}

We move on and discuss the NLO QCD correction to the aforementioned LO diagrams using a non-anticommutating $\gamma_5$ in HEFT where a similar phenomenon happens.
Let us first consider the NLO QCD corrections of the non-anomalous type to the LO diagrams of figure~\ref{dia:LO_HEFT} in HEFT. 
These corrections correspond to the two-loop diagrams of order $\alpha_s^3$ 
shown in figure~\ref{dia:nonsinglet_HEFT}. 
There are also a few  non-zero contributions of the anomalous type, i.e., non-vanishing diagrams involving quark triangles, at this perturbative order, 
shown in figure~\ref{dia:singlet_HEFT}, which we will discuss separately.
All the projectors used in computing these non-anomalous NLO QCD corrections are still the same as those used for the LO diagrams, 
and the only technical complication comes from performing the 
UV renormalization of the axial form factors regularized with a non-anticommuting $\gamma_5$. 
 The renormalization procedure described in section~\ref{sec:qqZHinHEFT_uv} is sufficient to renormalize the vector form factors  
 which is verified by checking the respective Ward identity.

We use now the prescription of refs.~\cite{Larin:1991tj,Larin:1993tq} for the axial vector current, sometimes referred to as Larin's prescription for short, and therefore the corresponding axial vector current renormalization constants in the ${\overline{\rm MS}}$ scheme get involved non-trivially at this order of perturbation theory, in addition to the QCD coupling and operator renormalization.  
For the non-anomalous NLO QCD corrections to the diagrams of figure~\ref{dia:LO_HEFT}, the normal form of the Ward identity should still hold for the axial vector current, exactly the same as for the vector counterpart.
However, to our surprise, if one just incorporates the usual UV counter-terms 
 arising from coupling constant, operator renormalization, and the compensation terms dictated by Larin's prescription, 
 the results for the axial form factors are still wrong, which manifests itself in the following checks. 
\begin{itemize}
\item 
The remaining pole structures in the axial form factors renormalized in this way do not match with
the prediction according to Catani's IR factorisation formula~\cite{Catani:1998bh}.

\item 
The $\epsilon^0$-order terms of the, renormalized and subtracted, vector and respective axial form factors differ, which subsequently 
implies the violation of the axial Ward identity for the process at hand.
\end{itemize}
The solution we found, which is now not hard to guess based on the above exposition of how 
to correct the LO result, is that one should also consistently compute 
the perturbative contributions to the extra local composite operator given in \eqref{eq:amendO}
that include both i) the perturbative expansion of $\mathrm{Z}_5^{h}(a_s)$ in $a_s$ and ii) 
the NLO (one-loop) corrections to this local composite operator (where the axial vector current involved is again treated by Larin's prescription). 
Following this line, we determine the expression for $\mathrm{Z}_{5,ns}^{h}(a_s)$
by imposing the equality between the finite remainders of the vector and (non-anomalous) axial form factors. 
We get
\begin{align} 
\label{eq:amend_Z5ns} 
\mathrm{Z}_{5, ns}^{h}(a_s) &=
1 \,+\,  a_s\, \left( 
\frac{-\beta_0}{\epsilon} 
\,+\, 
\frac{107}{18} C_A   
- 7 C_F - \frac{1}{9} n_f \right)
\,+\, \mathcal{O}(a^2_s) \, .
\end{align}
To summarize, the non-anomalous axial amplitude ${\cal A}_{axi(ns)}$ computed at two loops in HEFT 
 using a non-anticommuting $\gamma_5$ is renormalized according to\footnote{We remark that the particular organization 
 of the contributing terms as done in \eqref{eq:renorm-ns-amp} is made here for convenience. A systematic formulation of how the various pieces enter the computation can be made with an ``effective'' Lagrangian. This will be done at the end of this subsection.}
\begin{align}
\label{eq:renorm-ns-amp}
    {\cal A}^{\mu,\mathrm{NAC}}_{axi (ns)}(a_s) =Z^{ns}_{5,L}(a_s) Z^{ns}_{A,L}(a_s) Z_H(a_s) {\hat {\cal A}}^{\mu,\mathrm{NAC}}_{axi (ns)}({\hat a}_s) \,+\, 
    \mathcal{J}^{\mu,{\rm NAC}}_{ns} \, . 
\end{align}
In order to distinguish here bare and renormalized quantities we denote the unrenormalized amplitude  and QCD coupling with a hat.
The counterterm vertex in \eqref{eq:renorm-ns-amp} is given by
\begin{align}
\label{eq:amendo-ns}
     \mathcal{J}^{\mu,{\rm NAC}}_{ns}
    &=\mathrm{Z}_{5,ns}^{h}(a_s)\, \mathrm{\mathbf{C}} \, 
    \Big(\bar{v}(p_2) \,\left[\gamma^{\mu} \gamma_5\right]_{L} \, u(p_1) \Big) \nonumber\\
    &=\mathrm{Z}_{5,ns}^{h}(a_s) Z^{ns}_{5,L}(a_s) Z^{ns}_{A,L}(a_s)\, \mathrm{\mathbf{C}} \, 
    \Big(\bar{v}(p_2) \,\gamma^{\mu} \gamma_5 \, u(p_1) \Big)\,.
\end{align}
The constant $\mathrm{\mathbf{C}}$ is given below eq.~\eqref{eq:amendO}. 
The quantity $Z_H$ is the operator renormalization constant \eqref{eq:opZH} in the effective Lagrangian in HEFT; $Z^{ns}_{5,L}$ and $Z^{ns}_{A,L}$ 
are the renormalization constants for the non-singlet axial vector current in Larin's prescription~\cite{Larin:1991tj},
which we list here for convenience: 
\begin{align}
\label{eq:renorm-const-ns}
    &Z^{ns}_{A,L}=1+a_s^2 \frac{1}{\epsilon} \left( \frac{22}{3} C_F C_A -\frac{4}{3} C_F n_f\right)\,, \nonumber\\
    &Z^{ns}_{5,L}=1+a_s \left( -4 C_F\right)+a_s^2 \left( 22 C_F^2-\frac{107}{9} C_F C_A + \frac{2}{9} C_F n_f \right)\,.
\end{align}

If one does not invoke the Larin counterterms \eqref{eq:renorm-const-ns} then, of course, 
only the product $\mathrm{Z}_{5,ns}^{h,\mathrm{T}}(a_s) \equiv \mathrm{Z}_{5,ns}^{h}(a_s) Z^{ns}_{5,L}(a_s) Z^{ns}_{A,L}(a_s)$ 
as a whole can be determined in a chosen renormalization scheme (e.g.~in the $\overline{\mathrm{MS}}$ scheme used here) 
by demanding that the correct physical results (see above) are obtained.

\subsection*{UV renormalization of the anomalous diagrams in HEFT}
\label{suse:UVFSiC}

Regarding the anomalous diagrams at this perturbative order, shown by the non-vanishing ones in figure \ref{dia:singlet_HEFT}, a similar treatment as in the non-anomalous case can be applied,
albeit with a bit more complexity due to the ABJ anomaly~\cite{Adler:1969gk,Bell:1969ts}. 
Let us first discuss the UV-renormalized Ward identity that the non-vanishing four diagrams of figure~\ref{dia:singlet_HEFT}, where only the axial vector current contributes, 
must satisfy. 
Only the massless $b$-quark triangles make a non-zero contribution to the last four diagrams of figure~\ref{dia:singlet_HEFT}.
And those with $t$-quark loops are omitted as we work here in $n_f=5$ flavor HEFT. 
The contributions involve the coupling factor $g_{A,b} c_t$. 
The operator relation for the ABJ anomaly of the massless axial vector $b$-quark current $J_{5\mu}={\bar b}\gamma_\mu \gamma_5 b$
reads:
\begin{align}
\label{eq:ABJ}
    \Big[\partial^{\mu} J_{5\mu}\Big]_R=a_s\frac{1}{2}\Big[G\tilde G\Big]_R \, ,
\end{align}
where $G\tilde G =-\epsilon^{\mu\nu\rho\sigma}G^a_{\mu\nu}G^a_{\rho\sigma}$. 
The subscript $R$ indicates that these composite local operators need to be properly renormalized in order that this operator relation holds.
Let us denote the renormalized contribution to $\mathcal{A}_{axi (s)}$ as defined in \eqref{eq:Aaxisns} from the anomalous Feynman diagrams of figure~\ref{dia:singlet_HEFT} by 
\begin{equation}
\label{eq:axsheft}
 \mathcal{A}_{axi (s)} = \bar{v}(p_2) \, \mathbf{\Gamma}^{\mu}_{(s)} \, u(p_1)\, \varepsilon_{\mu}^{*} \, .
\end{equation}
Taking the divergence of $J_{5\mu}$ amounts to substituting $\varepsilon_{\mu}^{*} \to  q_{1,\mu}$.
By restoring the ABJ anomaly term on the right-hand side of \eqref{eq:consax} and repeating 
steps similar to eqs.~\eqref{eq:LSZ1} - \eqref{eq:pi-noeth}, we obtain for the anomalous
 contributions to $q\bar{q} \rightarrow ZH$ in HEFT the relation: 
\begin{equation}
  \bar{v}(p_2) \, \mathbf{\Gamma}^{\mu}_{(s)} \, u(p_1)\, q_{1,\mu} =
  \frac{a_s}{2} \big\langle H(q_2)\big| \big[G \tilde{G}(0)\big]_R \big| q(p_1) {\overline q}(p_2) \big\rangle \, ,
\label{eq:WardIABJ}
\end{equation}
where the kinematics of the matrix element on the r.h.s. obeys $p_1+p_2 -q_2=q_1$. As in the case of eq.~\eqref{eq:WardIH}
 there is an external momentum insertion $q^{\mu}_1$ by  the composite operator $\big[G \tilde{G}(0)\big]_R$.
Note that, at the two-loop order considered here, both sides of \eqref{eq:WardIABJ} are finite upon proper UV renormalization to be determined below.

\begin{figure}[htbp]
\begin{center}
\includegraphics[scale=0.45]{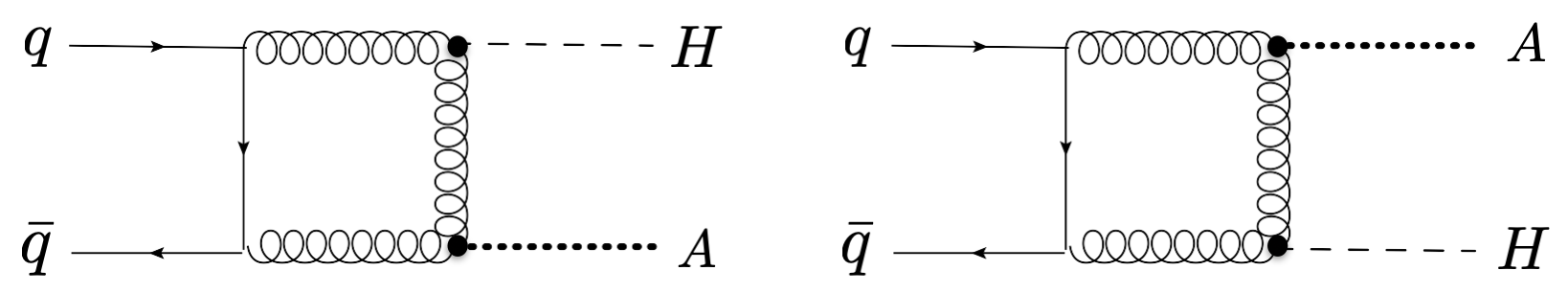}
\caption{Feynman diagrams contributing to the right-hand side of \eqref{eq:WardIABJ}. The blob associated with $H$ denotes the effective $Hgg$ vertex while the blob associated with the auxiliary pseudoscalar  $A$  results from the local operator $G\tilde G$.}
\label{dia:ABJ_RHS}
\end{center}
\end{figure}

The matrix element on the r.h.s. of \eqref{eq:WardIABJ} can be computed in perturbation theory order by order in $a_s$. 
The first non-vanishing term corresponds to the one-loop diagrams  depicted in figure~\ref{dia:ABJ_RHS}.

However, the relation \eqref{eq:WardIABJ} does not hold with the expressions we get for both sides. 
At the order we are considering, i.e.~NLO in $a_s$ w.r.t. the LO diagrams of figure \ref{dia:LO_HEFT}, none of the  flavor-singlet
axial current $Z_5$ factors of ref.~\cite{Larin:1993tq} (e.g.~listed in  eq.~(4.10) in ref.~\cite{Ahmed:2019udm}) contributes, 
because their non-vanishing perturbative terms start at order $a_s^2$ relative to the leading terms.
In view of our treatment of the non-anomalous contributions discussed previously, we therefore introduce for the renormalization
of the anomalous two-loop contributions  of figure~\ref{dia:singlet_HEFT} the local composite operator given in \eqref{eq:amendO} 
as an additional counterterm into the game, albeit with a new undetermined coefficient $\mathrm{Z}_{5, s}^{h}(a_s)$. 
The expression of $\mathrm{Z}_{5, s}^{h}(a_s)$ given in \eqref{eq:amend_Z5s} was determined such that the properly UV renormalized 
singlet axial current has a finite anomaly that does obey \eqref{eq:ABJ} and the Ward identity \eqref{eq:WardIABJ}.
With the concrete analytic expressions at hand, we obtain 
\begin{align} 
\label{eq:amend_Z5s} 
\mathrm{Z}_{5, s}^{h}(a_s) &=
1 \,+\,  a_s\, \left( 
-\frac{3}{2}\frac{1}{\epsilon} 
-\frac{3}{4} \right) 
\,+\, \mathcal{O}(a^2_s) \, .
\end{align}
We remark that the renormalization of the right-hand side of \eqref{eq:WardIABJ} at the perturbative order in question involves the following mixed counterterm: 
\begin{align}
\label{eq:amend_ZGJ}
\mathrm{Z}_{GJ}^{h}(a_s)\,  
\big(\bar{v}(p_2) \,\gamma_{\mu} \gamma_5 \, u(p_1) \, q^{\mu}_{1} \big)      
\end{align}
where $\mathrm{Z}_{GJ}^{h}(a_s) = a_s\, \left(\frac{24 } {\epsilon} C_F\right) + \mathcal{O}(a_s^2)$ 
is determined by the requirement of minimally subtracting all poles of the Feynman diagrams in figure \ref{dia:ABJ_RHS} (under the convention of setting the coupling factors associated with the effective $Hgg$ vertex and the local operator $G\tilde{G}$ to be one).
~\\

To summarize, our explicit computation of the class-I contributions to $q\bar{q} \rightarrow ZH$ and their QCD corrections in HEFT, as presented in the preceding sections, shows that if one uses a non-anticommuting $\gamma_5$ in dimensional regularization one  has to introduce an 
additional counterterm 
$\mathrm{Z}_5^{h}(a_s)\, \mathrm{\mathbf{C}} \, \Big(\bar{q}_R \,\left[\gamma^{\mu} \gamma_5\right]_{L} \, q_R\Big) Z_{\mu}\, H$ 
into the final properly renormalized effective Lagrangian. 
To be more specific, the complete form of the properly renormalized effective Lagrangian with a non-anticommuting $\gamma_5$ 
that one should use in computing the QCD corrections to $\mathcal{O}(\alpha_s^3)$  to the class-I 
contributions to $q\bar{q} \rightarrow ZH$ in HEFT reads as 
\begin{align}
\label{eq:Rlagrangian}
\mathcal{L}_{R} = \Big[\mathcal{L}_{c}+\mathcal{L}_{\mathrm{heff}}\Big]_{R} + 
\kappa~\mathrm{Z}_5^{h}(a_s)\, \mathrm{\mathbf{C}} \, \Big(\bar{q}_R(x) \,\left[\gamma^{\mu} \gamma_5\right]_{L} \, q_R(x)\Big) Z_{\mu}(x)\, H(x)\,,
\end{align}
where $\Big[\mathcal{L}_{c}+\mathcal{L}_{\mathrm{heff}}\Big]_{R}$ denotes the renormalized form of 
$\mathcal{L}_c + \mathcal{L}_{\mathrm{heff}}$ given by \eqref{eq:lagbhz} and \eqref{eq:HEFT-Lag}.
The explicit perturbative expressions for $\mathrm{Z}_{5, ns}^{h}(a_s)$ and $\mathrm{Z}_{5, s}^{h}(a_s)$ were determined in \eqref{eq:amend_Z5ns} and \eqref{eq:amend_Z5s} to the first order in $a_s$.
Furthermore, $\kappa_{ns} = c_t g_{A,q}$ and  $\kappa_{s} = c_t g_{A,b}$.
~\\

We conclude this subsection with a short comment on another class of non-vanishing top-loop contributions to the amplitude of $q{\bar q} \to Z H$ proportional to $\lambda_t$. 
In these contributions, which are called class-II in ref.~\cite{Brein:2011vx} and start at two loops, i.e. order $a_s^2$, the $Z$ boson couples to a virtual top-quark loop, from which the Higgs boson is also radiated. 
Charge conjugation invariance  dictates that only the axial vector current contributes at order $a_s^2$. Thus these contributions are proportional 
to the product of couplings $\lambda_t \, g_{A,t}$. It was shown in ref.~\cite{Brein:2011vx}, where the axial vector current was regularized according to Larin's prescription, that in the limit $m_t\to \infty$ the non-vanishing part of the class-II contributions involves only one structure, namely ${\bar q}\gamma_\mu \gamma_5 q Z^\mu H$. 
The investigation of the NLO QCD corrections 
to these two-loop class-II contributions in the limit $m_t\to\infty$ is beyond the scope of this work and will be deferred to a future investigation.


\subsection{The class-I Feynman diagrams at two loops without taking the heavy-top limit}
\label{sec:result_withoutHT}

Our investigations above of the class-I type contributions with HEFT show that correct results (i.e., results that obey chiral invariance and the
correct  Ward identity) are obtained 
for both the vector part and the (non-anomalous) axial part with an anticommuting $\gamma_5$. 
But when employing a non-anticommuting $\gamma_5$ in the computation of the same set of 
(non-anomalous) Feynman diagrams in HEFT, it seems that there are some missing pieces, 
which calls for additional amendment terms as explicitly determined above.
In order to have a better understanding of this issue,  
we would like to know how these class-I Feynman diagrams behave in the full six-flavor
theory without taking heavy-top limit.
In particular, we want to check whether the equality \eqref{eq:FFD_vEa_AC5} holds 
 in a computation with a non-anticommuting $\gamma_5$, which then implies whether the
Ward identity \eqref{eq:WI_AC5} is satisfied without the need of additional amendment terms (at two-loop order).

There are in total 6 Feynman diagrams for the class-I type contributions at two-loop order with a finite top-quark mass, with samples depicted in figure~\ref{dia:LO_full}. 
We generate the (unreduced) symbolic expressions using an extension of \gosam~\cite{Cullen:2014yla,Jones:2016bci}. 
After applying the integration-by-parts (IBP)~\cite{Tkachov:1981wb,Chetyrkin:1981qh} relations obtained with \kira~\cite{Maierhofer:2018gpa}, they are reduced into linear combinations of 55 master loop integrals.
The hardest one is a 7-propagator loop integral that depends on 5 scales (2 Mandelstam variables and 3 physical masses), corresponding 
to the topology of the first diagram in figure~\ref{dia:LO_full}, for which no analytic results are known yet.
For our purpose of a numerical verification of \eqref{eq:FFD_vEa_AC5}, all 55 master integrals are
computed using {\tt pySecdec}~\cite{Borowka:2017idc} at one chosen kinematic point.

Both the vector and axial vector form factors of the two-loop class-I diagrams in the full theory, defined in complete analogy 
to \eqref{eq:FFdecomp-qqZH-vec} and \eqref{eq:FFD_axi}, respectively, and projected out using the same projectors as employed in the previous computations done in HEFT, 
are finite without renormalization or infrared subtraction, 
because they are the Born level contributions of this type. 
However, the master integrals involved could and indeed contain spurious poles which all cancel in their final linear combinations making up the form factors. 
In this numerical check, we tried three different sets of master bases: i) loop integrals without irreducible numerators,
ii) loop integrals where irreducible numerators are favoured over denominators raised to powers,
and also iii) quasi-finite 
loop integrals~\cite{vonManteuffel:2014qoa}, each of which is not unique in general and partially 
depends on the integral-ordering in use.
Concerning the particular choice we take, the quasi-finite master basis exhibits the least spurious poles,
albeit still starting from 
$1/\epsilon^2$, while the master basis with numerators performs the worst in the numerical evaluation 
with {\tt pySecdec}~\cite{Borowka:2017idc}.
We note that the master basis without irreducible numerators determined by \kira~\cite{Maierhofer:2018gpa} for our integral 
family  exhibits automatically the feature where their rational coefficients in the IBP table, and hence the reduced amplitudes, 
have their denominators' D-dependence factorized from the external kinematics~\cite{Smirnov:2020quc,Usovitsch:2020jrk}.
Furthermore, we observe that vector and axial form factors not only share exactly 
the same master basis, but also their respective rational 
coefficients agree, albeit, only to the leading term in their Laurent expansions
around D=4 dimensions\footnote{If the ``basis of loop integrals" 
in use is not truly minimal, or there are hidden zeros among them, then this structure is not necessarily observed.}, 
just as in their one-loop counterparts in HEFT discussed at the end of section~\ref{sec:result_asitis}.
Under this condition, if one is only interested in checking the difference between these two sets of form factors, 
then  all pieces needed to this end are in fact those that are used for demonstrating
the cancellation of all spurious poles in these finite form factors. 
This property can be exploited to greatly improve the level of numerical accuracy 
of the comparison without increasing the 
computational cost, as the most complicated parts of master integrals required to obtain the form factors or amplitudes to 
$\epsilon^0$ are not needed for this purpose. 
Eventually, we obtain agreement between the vector and axial form factors of the two-loop class-I diagrams 
in the full six-flavor theory (i.e., without the effective Higgs-gluon vertex) at the chosen kinematic point within the numerical uncertainty.
In particular, taking advantage of the aforementioned insight, the suspicious 4-th vector and axial vector form factor agree
with each other to the fourth significant digit, which we deem to be quite sufficient for our 
purpose\footnote{The two numbers for this comparison are obtained after running {\tt pySecdec} for about 10 hours on a desk-top 
computer with 6 CPUs using the Vegas integrator~\cite{Hahn:2004fe}, whose errors are estimated conservatively by 
the addition-in-quadrature formula to be a few per mille.
If one compares at the level of the complete finite $\epsilon^0$-order, the numerical uncertainty will be roughly 10 times 
larger and the agreement can be seen only in the first 2 or 3 significant digits.}.
~\\

With the outcome of this critical check we conclude the following. 
If one naively computes the class-I diagrams in HEFT with a non-anticommuting $\gamma_5$, 
some terms are missing, namely terms that involve the effective vertex $q\gamma_\mu \gamma_5 \bar{q}Z^{\mu}H$.
We have restored and determined these additional counterterms by enforcing the respective  Ward identities as 
discussed in section~\ref{sec:result_asitis} and~\ref{sec:result_asitisNLO}.
Computing analytically this particular missing piece \textit{directly} by applying the heavy top-mass expansion 
procedure to these two-loop class-I diagrams and confirming the result deduced above would also be an interesting thing to do, which we however defer to future work.

Our investigations conducted above show that the presence or absence of diagrams with the effective vertex  $q\gamma_\mu \gamma_5 \bar{q}Z^{\mu}H$ 
in the heavy top-mass expansion of the class-I
diagrams depends on the specific $\gamma_5$ prescription in use.  
When the axial vector current is regularized using a non-anticommuting $\gamma_5$ these are apparently needed.
 This observation further strengthens the common lore that one should be very careful with taking claims (or assuming conditions) 
 which were established with an anticommuting $\gamma_5$ in a computation where a non-anticommuting $\gamma_5$ is employed instead in dimensional regularization.

\section{Conclusions}
\label{sec:conc}

In the first part of this article (sections~\ref{sec:AfromV} and~\ref{sec:WI}), we considered first  the vector current and non-anomalous axial vector current amplitudes
$b \bar{b} \rightarrow Z h$ proportional to the $b$-quark Yukawa coupling at two loops for a CP-even and CP-odd Higgs boson $h = H, A$,
 respectively. We showed that these polarized amplitudes
can be obtained, when the $b$ quark is taken to be kinematically massless, solely from  the vector form factors of properly grouped classes of diagrams for $ZH$ production. 
 The computation of these form factors does not involve 
the axial vector current and hence $\gamma_5$. 
Subsequently, we have compared with the previous results on $b \bar{b} \rightarrow Z H$ of ref.~\cite{Ahmed:2019udm} 
to two-loop order (where different projectors were used with a non-anticommuting $\gamma_5$). As expected, agreement
of the axial part of the amplitudes can only be obtained at the level of properly defined finite remainders in four dimensions.
Furthermore, the Ward identities for the QCD corrections to these $b$-quark Yukawa coupling-dependent contributions 
to $b \bar{b} \rightarrow Z h$, $h=H,A$, are derived and verified explicitly in section~\ref{sec:WI}.

In addition, we determined the two-loop contributions 
to $b \bar{b} \rightarrow Z h$ corresponding to diagrams that involve $b$- and $t$-quark triangles and the axial vector current
in the heavy-top limit (cf. figure~\ref{dia:singlet}). For $m_t\to\infty$ the top-loop induced triangle diagrams are not vanishing. The explicit dependence on the renormalization scale  $\mu_R$ cancels in the total two-loop triangle contribution. 
~\\

In the second part of this article (sections~\ref{sec:qqZHinHEFT} and~\ref{sec:G5pitfall}), we considered a class of top-Yukawa coupling dependent contributions
 to the amplitude of $q\bar{q} \rightarrow ZH$, namely the so-called class-I diagrams. Here the Higgs boson is radiated from the internal top-quark loop while the $Z$ boson is emitted from the external light quark line. We computed these contributions to  
  $\mathcal{O}(\alpha_s^3)$ in the heavy-top limit using the Higgs-gluon effective Lagrangian (HEFT). 
 We obtained the analytical expressions of the UV renormalized vector form factors to two-loop order and verified 
their infrared poles by comparing to Catani's infrared factorization formula.
For computing the axial vector form factors of the non-anomalous diagrams, an anticommuting $\gamma_5$ can be employed, which results in exactly the same UV renormalized expressions as their vector counterparts.

In an attempt to re-compute the QCD corrections to the same class-I Feynman diagrams in HEFT, but with a non-anticommuting $\gamma_5$,
a technical pitfall was noticed and discussed in detail in section~\ref{sec:G5pitfall}. 
One may look at this issue from two perspectives.
If one limits the scope to be within HEFT, then there are some additional local composite operators that one has to include 
when using a non-anticommuting $\gamma_5$ in the computation of class-I contributions to $q\bar{q} \rightarrow ZH$, 
as summarized in \eqref{eq:Rlagrangian}. 
 On the other hand, if one looks at it from the point of view of the original full six-flavor 
 theory, then our analysis in section~\ref{sec:result_withoutHT} 
implies the following.  If a non-anticommuting $\gamma_5$ is used in the axial vector current, then in the  infinite top-mass limit
certain heavy-mass expanded diagrams survive  that are absent in a respective computation where 
an anticommuting $\gamma_5$ is used. 
Therefore, our results show that the presence or absence of certain heavy-mass expanded diagrams in the infinite-mass limit of a scattering amplitude with an axial vector current actually depends on the particular $\gamma_5$  prescription in use.


\section*{Acknowledgements}

We thank V.~Ravindran for comments on the manuscript.
L.C. is grateful to R.~Harlander and M.~Niggetiedt for a number of clarifying discussions regarding the heavy mass expansion and also for their hospitality during a visit to Aachen.
The work of T.A. received funding from the European Research Council (ERC) under the European Union’s Horizon 2020 research 
and innovation programme, \textit{Novel structures in scattering amplitudes} (grant agreement No. 725110). 
The work of M.C. was supported by the Deutsche Forschungsgemeinschaft under grant 396021762 -- TRR 257.
The figures of the Feynman diagrams are generated using \tikz~\cite{Ellis:2016jkw} and  \xml~\cite{nicolas_deutschmann_2016_164393}. 
We have employed \qgraf~\cite{Nogueira:1991ex}, \form~\cite{Vermaseren:2000nd}, \litered~\cite{Lee:2012cn}, \reduze~\cite{Studerus:2009ye,vonManteuffel:2012np} and \fermat~\cite{fermat} in various parts of the computation.

\bibliography{qqzh} 

\providecommand{\href}[2]{#2}\begingroup\raggedright\begin{thebibliography}{10}

\bibitem{Aaboud:2018zhk}
{\bfseries ATLAS} Collaboration, M.~Aaboud {\em et~al.}, {\it {Observation of
  $H \rightarrow b\bar{b}$ decays and $VH$ production with the ATLAS
  detector}},  \href{http://dx.doi.org/10.1016/j.physletb.2018.09.013}{{\em
  Phys. Lett.} {\bfseries B786} (2018) 59--86},
\href{http://arxiv.org/abs/1808.08238}{{\ttfamily arXiv:1808.08238 [hep-ex]}}.

\bibitem{Sirunyan:2018kst}
{\bfseries CMS} Collaboration, A.~M. Sirunyan {\em et~al.}, {\it {Observation
  of Higgs boson decay to bottom quarks}},
  \href{http://dx.doi.org/10.1103/PhysRevLett.121.121801}{{\em Phys. Rev.
  Lett.} {\bfseries 121} no.~12, (2018) 121801},
\href{http://arxiv.org/abs/1808.08242}{{\ttfamily arXiv:1808.08242 [hep-ex]}}.

\bibitem{Butterworth:2008iy}
J.~M. Butterworth, A.~R. Davison, M.~Rubin, and G.~P. Salam, {\it {Jet
  substructure as a new Higgs search channel at the LHC}},
  \href{http://dx.doi.org/10.1103/PhysRevLett.100.242001}{{\em Phys. Rev.
  Lett.} {\bfseries 100} (2008) 242001},
\href{http://arxiv.org/abs/0802.2470}{{\ttfamily arXiv:0802.2470 [hep-ph]}}.

\bibitem{Brein:2003wg}
O.~Brein, A.~Djouadi, and R.~Harlander, {\it {NNLO QCD corrections to the
  Higgs-strahlung processes at hadron colliders}},
  \href{http://dx.doi.org/10.1016/j.physletb.2003.10.112}{{\em Phys. Lett.}
  {\bfseries B579} (2004) 149--156},
\href{http://arxiv.org/abs/hep-ph/0307206}{{\ttfamily arXiv:hep-ph/0307206
  [hep-ph]}}.

\bibitem{Brein:2011vx}
O.~Brein, R.~Harlander, M.~Wiesemann, and T.~Zirke, {\it {Top-Quark Mediated
  Effects in Hadronic Higgs-Strahlung}},
  \href{http://dx.doi.org/10.1140/epjc/s10052-012-1868-6}{{\em Eur. Phys. J.}
  {\bfseries C72} (2012) 1868},
\href{http://arxiv.org/abs/1111.0761}{{\ttfamily arXiv:1111.0761 [hep-ph]}}.

\bibitem{Ferrera:2014lca}
G.~Ferrera, M.~Grazzini, and F.~Tramontano, {\it {Associated ZH production at
  hadron colliders: the fully differential NNLO QCD calculation}},
  \href{http://dx.doi.org/10.1016/j.physletb.2014.11.040}{{\em Phys. Lett.}
  {\bfseries B740} (2015) 51--55},
\href{http://arxiv.org/abs/1407.4747}{{\ttfamily arXiv:1407.4747 [hep-ph]}}.

\bibitem{Ahmed:2014cla}
T.~Ahmed, M.~Mahakhud, N.~Rana, and V.~Ravindran, {\it {Drell-Yan Production at
  Threshold to Third Order in QCD}},
  \href{http://dx.doi.org/10.1103/PhysRevLett.113.112002}{{\em Phys. Rev.
  Lett.} {\bfseries 113} no.~11, (2014) 112002},
\href{http://arxiv.org/abs/1404.0366}{{\ttfamily arXiv:1404.0366 [hep-ph]}}.

\bibitem{Li:2014bfa}
Y.~Li, A.~von Manteuffel, R.~M. Schabinger, and H.~X. Zhu, {\it {N$^3$LO Higgs
  boson and Drell-Yan production at threshold: The one-loop two-emission
  contribution}},  \href{http://dx.doi.org/10.1103/PhysRevD.90.053006}{{\em
  Phys. Rev.} {\bfseries D90} no.~5, (2014) 053006},
\href{http://arxiv.org/abs/1404.5839}{{\ttfamily arXiv:1404.5839 [hep-ph]}}.

\bibitem{Catani:2014uta}
S.~Catani, L.~Cieri, D.~de~Florian, G.~Ferrera, and M.~Grazzini, {\it
  {Threshold resummation at N$^3$LL accuracy and soft-virtual cross sections at
  N$^3$LO}},  \href{http://dx.doi.org/10.1016/j.nuclphysb.2014.09.012}{{\em
  Nucl. Phys.} {\bfseries B888} (2014) 75--91},
\href{http://arxiv.org/abs/1405.4827}{{\ttfamily arXiv:1405.4827 [hep-ph]}}.

\bibitem{Kumar:2014uwa}
M.~C. Kumar, M.~K. Mandal, and V.~Ravindran, {\it {Associated production of
  Higgs boson with vector boson at threshold N$^{3}$LO in QCD}},
  \href{http://dx.doi.org/10.1007/JHEP03(2015)037}{{\em JHEP} {\bfseries 03}
  (2015) 037},
\href{http://arxiv.org/abs/1412.3357}{{\ttfamily arXiv:1412.3357 [hep-ph]}}.

\bibitem{Campbell:2016jau}
J.~M. Campbell, R.~K. Ellis, and C.~Williams, {\it {Associated production of a
  Higgs boson at NNLO}},  \href{http://dx.doi.org/10.1007/JHEP06(2016)179}{{\em
  JHEP} {\bfseries 06} (2016) 179},
\href{http://arxiv.org/abs/1601.00658}{{\ttfamily arXiv:1601.00658 [hep-ph]}}.

\bibitem{Ferrera:2017zex}
G.~Ferrera, G.~Somogyi, and F.~Tramontano, {\it {Associated production of a
  Higgs boson decaying into bottom quarks at the LHC in full NNLO QCD}},
  \href{http://dx.doi.org/10.1016/j.physletb.2018.03.021}{{\em Phys. Lett.}
  {\bfseries B780} (2018) 346--351},
\href{http://arxiv.org/abs/1705.10304}{{\ttfamily arXiv:1705.10304 [hep-ph]}}.

\bibitem{Ahmed:2019udm}
T.~Ahmed, A.~H. Ajjath, L.~Chen, P.~K. Dhani, P.~Mukherjee, and V.~Ravindran,
  {\it {Polarised Amplitudes and Soft-Virtual Cross Sections for $b\bar b
  \rightarrow ZH$ at NNLO in QCD}},
  \href{http://dx.doi.org/10.1007/JHEP01(2020)030}{{\em JHEP} {\bfseries 01}
  (2020) 030},
\href{http://arxiv.org/abs/1910.06347}{{\ttfamily arXiv:1910.06347 [hep-ph]}}.

\bibitem{Chen:2019wyb}
L.~Chen, {\it {A prescription for projectors to compute helicity amplitudes in
  D dimensions}},
\href{http://arxiv.org/abs/1904.00705}{{\ttfamily arXiv:1904.00705 [hep-ph]}}.

\bibitem{Larin:1991tj}
S.~A. Larin and J.~A.~M. Vermaseren, {\it {The $\alpha_s^3$ corrections to the
  Bjorken sum rule for polarized electroproduction and to the Gross-Llewellyn
  Smith sum rule}},
\href{http://dx.doi.org/10.1016/0370-2693(91)90839-I}{{\em Phys. Lett.}
  {\bfseries B259} (1991) 345--352}.

\bibitem{Larin:1993tq}
S.~A. Larin, {\it {The Renormalization of the axial anomaly in dimensional
  regularization}},  \href{http://dx.doi.org/10.1016/0370-2693(93)90053-K}{{\em
  Phys. Lett.} {\bfseries B303} (1993) 113--118},
\href{http://arxiv.org/abs/hep-ph/9302240}{{\ttfamily arXiv:hep-ph/9302240
  [hep-ph]}}.

\bibitem{Bardeen:1972vi}
W.~A. Bardeen, R.~Gastmans, and B.~E. Lautrup, {\it {Static quantities in
  Weinberg's model of weak and electromagnetic interactions}},
\href{http://dx.doi.org/10.1016/0550-3213(72)90218-0}{{\em Nucl. Phys.}
  {\bfseries B46} (1972) 319--331}.

\bibitem{Chanowitz:1979zu}
M.~S. Chanowitz, M.~Furman, and I.~Hinchliffe, {\it {The Axial Current in
  Dimensional Regularization}},
\href{http://dx.doi.org/10.1016/0550-3213(79)90333-X}{{\em Nucl. Phys.}
  {\bfseries B159} (1979) 225--243}.

\bibitem{Gottlieb:1979ix}
S.~A. Gottlieb and J.~T. Donohue, {\it {The Axial Vector Current and
  Dimensional Regularization}},
\href{http://dx.doi.org/10.1103/PhysRevD.20.3378}{{\em Phys. Rev.} {\bfseries
  D20} (1979) 3378}.

\bibitem{Ovrut:1981ne}
B.~A. Ovrut, {\it {Axial Vector Ward Identities and Dimensional
  Regularization}},
\href{http://dx.doi.org/10.1016/0550-3213(83)90511-4}{{\em Nucl. Phys.}
  {\bfseries B213} (1983) 241--265}.

\bibitem{Espriu:1982bw}
D.~Espriu and R.~Tarrach, {\it {Renormalization of the Axial Anomaly
  Operators}},
\href{http://dx.doi.org/10.1007/BF01573750}{{\em Z. Phys.} {\bfseries C16}
  (1982) 77}.

\bibitem{Korner:1991sx}
J.~G. Korner, D.~Kreimer, and K.~Schilcher, {\it {A Practicable gamma(5) scheme
  in dimensional regularization}},
\href{http://dx.doi.org/10.1007/BF01559471}{{\em Z. Phys.} {\bfseries C54}
  (1992) 503--512}.

\bibitem{Adler:1969gk}
S.~L. Adler, {\it {Axial vector vertex in spinor electrodynamics}},
  \href{http://dx.doi.org/10.1103/PhysRev.177.2426}{{\em Phys. Rev.} {\bfseries
  177} (1969) 2426--2438}.
[,241(1969)].

\bibitem{Bell:1969ts}
J.~S. Bell and R.~Jackiw, {\it {A PCAC puzzle: $\pi^0 \to \gamma \gamma$ in the
  $\sigma$ model}},
\href{http://dx.doi.org/10.1007/BF02823296}{{\em Nuovo Cim.} {\bfseries A60}
  (1969) 47--61}.

\bibitem{Smirnov:2002pj}
V.~A. Smirnov, {\it {Applied asymptotic expansions in momenta and masses}},
{\em Springer Tracts Mod. Phys.} {\bfseries 177} (2002) 1--262.

\bibitem{Patel:2015tea}
H.~H. Patel, {\it {Package-X: A Mathematica package for the analytic
  calculation of one-loop integrals}},
  \href{http://dx.doi.org/10.1016/j.cpc.2015.08.017}{{\em Comput. Phys.
  Commun.} {\bfseries 197} (2015) 276--290},
\href{http://arxiv.org/abs/1503.01469}{{\ttfamily arXiv:1503.01469 [hep-ph]}}.

\bibitem{Bernreuther:2005rw}
W.~Bernreuther, R.~Bonciani, T.~Gehrmann, R.~Heinesch, T.~Leineweber, and
  E.~Remiddi, {\it {Two-loop QCD corrections to the heavy quark form-factors:
  Anomaly contributions}},
  \href{http://dx.doi.org/10.1016/j.nuclphysb.2005.06.025}{{\em Nucl. Phys.}
  {\bfseries B723} (2005) 91--116},
\href{http://arxiv.org/abs/hep-ph/0504190}{{\ttfamily arXiv:hep-ph/0504190
  [hep-ph]}}.

\bibitem{Chetyrkin:1998mw}
K.~G. Chetyrkin, B.~A. Kniehl, M.~Steinhauser, and W.~A. Bardeen, {\it
  {Effective QCD interactions of CP odd Higgs bosons at three loops}},
  \href{http://dx.doi.org/10.1016/S0550-3213(98)00594-X}{{\em Nucl. Phys.}
  {\bfseries B535} (1998) 3--18},
\href{http://arxiv.org/abs/hep-ph/9807241}{{\ttfamily arXiv:hep-ph/9807241
  [hep-ph]}}.

\bibitem{Lehmann:1954rq}
H.~Lehmann, K.~Symanzik, and W.~Zimmermann, {\it {On the formulation of
  quantized field theories}},
\href{http://dx.doi.org/10.1007/BF02731765}{{\em Nuovo Cim.} {\bfseries 1}
  (1955) 205--225}.

\bibitem{Kniehl:1995tn}
B.~A. Kniehl and M.~Spira, {\it {Low-energy theorems in Higgs physics}},
  \href{http://dx.doi.org/10.1007/s002880050007}{{\em Z. Phys.} {\bfseries C69}
  (1995) 77--88},
\href{http://arxiv.org/abs/hep-ph/9505225}{{\ttfamily arXiv:hep-ph/9505225
  [hep-ph]}}.

\bibitem{Kramer:1996iq}
M.~Kramer, E.~Laenen, and M.~Spira, {\it {Soft gluon radiation in Higgs boson
  production at the LHC}},
  \href{http://dx.doi.org/10.1016/S0550-3213(97)00679-2}{{\em Nucl. Phys.}
  {\bfseries B511} (1998) 523--549},
\href{http://arxiv.org/abs/hep-ph/9611272}{{\ttfamily arXiv:hep-ph/9611272
  [hep-ph]}}.

\bibitem{Chetyrkin:1997iv}
K.~G. Chetyrkin, B.~A. Kniehl, and M.~Steinhauser, {\it {Hadronic Higgs decay
  to order alpha-s**4}},
  \href{http://dx.doi.org/10.1103/PhysRevLett.79.353}{{\em Phys. Rev. Lett.}
  {\bfseries 79} (1997) 353--356},
\href{http://arxiv.org/abs/hep-ph/9705240}{{\ttfamily arXiv:hep-ph/9705240
  [hep-ph]}}.

\bibitem{Nielsen:1975ph}
N.~K. Nielsen, {\it {Gauge Invariance and Broken Conformal Symmetry}},
\href{http://dx.doi.org/10.1016/0550-3213(75)90378-8}{{\em Nucl. Phys.}
  {\bfseries B97} (1975) 527--540}.

\bibitem{Spiridonov:1988md}
V.~P. Spiridonov and K.~G. Chetyrkin, {\it {Nonleading mass corrections and
  renormalization of the operators m psi-bar psi and g**2(mu nu)}},  {\em Sov.
  J. Nucl. Phys.} {\bfseries 47} (1988) 522--527.
[Yad. Fiz.47,818(1988)].

\bibitem{Kataev:1981gr}
A.~L. Kataev, N.~V. Krasnikov, and A.~A. Pivovarov, {\it {Two Loop Calculations
  for the Propagators of Gluonic Currents}},
  \href{http://dx.doi.org/10.1016/0550-3213(82)90338-8,
  10.1016/S0550-3213(97)00101-6}{{\em Nucl. Phys.} {\bfseries B198} (1982)
  508--518}, \href{http://arxiv.org/abs/hep-ph/9612326}{{\ttfamily
  arXiv:hep-ph/9612326 [hep-ph]}}.
[Erratum: Nucl. Phys.B490,505(1997)].

\bibitem{Gehrmann:2015ora}
T.~Gehrmann, A.~von Manteuffel, and L.~Tancredi, {\it {The two-loop helicity
  amplitudes for $ q\overline{q}^{\prime}\to {V}_1{V}_2\to 4 $ leptons}},
  \href{http://dx.doi.org/10.1007/JHEP09(2015)128}{{\em JHEP} {\bfseries 09}
  (2015) 128},
\href{http://arxiv.org/abs/1503.04812}{{\ttfamily arXiv:1503.04812 [hep-ph]}}.

\bibitem{hepforge}
\url{https://vvamp.hepforge.org/}.

\bibitem{Catani:1998bh}
S.~Catani, {\it {The Singular behavior of QCD amplitudes at two loop order}},
  \href{http://dx.doi.org/10.1016/S0370-2693(98)00332-3}{{\em Phys. Lett.}
  {\bfseries B427} (1998) 161--171},
\href{http://arxiv.org/abs/hep-ph/9802439}{{\ttfamily arXiv:hep-ph/9802439
  [hep-ph]}}.

\bibitem{Sterman:2002qn}
G.~F. Sterman and M.~E. Tejeda-Yeomans, {\it {Multiloop amplitudes and
  resummation}},  \href{http://dx.doi.org/10.1016/S0370-2693(02)03100-3}{{\em
  Phys. Lett.} {\bfseries B552} (2003) 48--56},
\href{http://arxiv.org/abs/hep-ph/0210130}{{\ttfamily arXiv:hep-ph/0210130
  [hep-ph]}}.

\bibitem{Becher:2009cu}
T.~Becher and M.~Neubert, {\it {Infrared singularities of scattering amplitudes
  in perturbative QCD}},
  \href{http://dx.doi.org/10.1103/PhysRevLett.102.162001,
  10.1103/PhysRevLett.111.199905}{{\em Phys. Rev. Lett.} {\bfseries 102} (2009)
  162001}, \href{http://arxiv.org/abs/0901.0722}{{\ttfamily arXiv:0901.0722
  [hep-ph]}}.
[Erratum: Phys. Rev. Lett.111,no.19,199905(2013)].

\bibitem{Becher:2009qa}
T.~Becher and M.~Neubert, {\it {On the Structure of Infrared Singularities of
  Gauge-Theory Amplitudes}},
  \href{http://dx.doi.org/10.1088/1126-6708/2009/06/081,
  10.1007/JHEP11(2013)024}{{\em JHEP} {\bfseries 06} (2009) 081},
  \href{http://arxiv.org/abs/0903.1126}{{\ttfamily arXiv:0903.1126 [hep-ph]}}.
[Erratum: JHEP11,024(2013)].

\bibitem{Gardi:2009qi}
E.~Gardi and L.~Magnea, {\it {Factorization constraints for soft anomalous
  dimensions in QCD scattering amplitudes}},
  \href{http://dx.doi.org/10.1088/1126-6708/2009/03/079}{{\em JHEP} {\bfseries
  03} (2009) 079},
\href{http://arxiv.org/abs/0901.1091}{{\ttfamily arXiv:0901.1091 [hep-ph]}}.

\bibitem{tHooft:1972tcz}
G.~'t~Hooft and M.~J.~G. Veltman, {\it {Regularization and Renormalization of
  Gauge Fields}},
\href{http://dx.doi.org/10.1016/0550-3213(72)90279-9}{{\em Nucl. Phys.}
  {\bfseries B44} (1972) 189--213}.

\bibitem{Breitenlohner:1977hr}
P.~Breitenlohner and D.~Maison, {\it {Dimensional Renormalization and the
  Action Principle}},
\href{http://dx.doi.org/10.1007/BF01609069}{{\em Commun. Math. Phys.}
  {\bfseries 52} (1977) 11--38}.

\bibitem{Akyeampong:1973xi}
D.~A. Akyeampong and R.~Delbourgo, {\it {Dimensional regularization, abnormal
  amplitudes and anomalies}},
\href{http://dx.doi.org/10.1007/BF02786835}{{\em Nuovo Cim.} {\bfseries A17}
  (1973) 578--586}.

\bibitem{Zijlstra:1992kj}
E.~B. Zijlstra and W.~L. van Neerven, {\it {${\cal O}( \alpha_s^2)$ correction
  to the structure function $F_3$ ($x, Q^2$) in deep inelastic neutrino -
  hadron scattering}},
\href{http://dx.doi.org/10.1016/0370-2693(92)91277-G}{{\em Phys. Lett.}
  {\bfseries B297} (1992) 377--384}.

\bibitem{Cullen:2014yla}
G.~Cullen {\em et~al.}, {\it {G$\scriptsize{O}$S$\scriptsize{AM}$-2.0: a tool
  for automated one-loop calculations within the Standard Model and beyond}},
  \href{http://dx.doi.org/10.1140/epjc/s10052-014-3001-5}{{\em Eur. Phys. J.}
  {\bfseries C74} no.~8, (2014) 3001},
\href{http://arxiv.org/abs/1404.7096}{{\ttfamily arXiv:1404.7096 [hep-ph]}}.

\bibitem{Jones:2016bci}
S.~P. Jones, {\it {Automation of 2-loop Amplitude Calculations}},
  \href{http://dx.doi.org/10.22323/1.260.0069}{{\em PoS} {\bfseries LL2016}
  (2016) 069},
\href{http://arxiv.org/abs/1608.03846}{{\ttfamily arXiv:1608.03846 [hep-ph]}}.

\bibitem{Tkachov:1981wb}
F.~V. Tkachov, {\it {A Theorem on Analytical Calculability of Four Loop
  Renormalization Group Functions}},
\href{http://dx.doi.org/10.1016/0370-2693(81)90288-4}{{\em Phys. Lett.}
  {\bfseries 100B} (1981) 65--68}.

\bibitem{Chetyrkin:1981qh}
K.~G. Chetyrkin and F.~V. Tkachov, {\it {Integration by Parts: The Algorithm to
  Calculate beta Functions in 4 Loops}},
\href{http://dx.doi.org/10.1016/0550-3213(81)90199-1}{{\em Nucl. Phys.}
  {\bfseries B192} (1981) 159--204}.

\bibitem{Maierhofer:2018gpa}
P.~Maierhöfer and J.~Usovitsch, {\it {Kira 1.2 Release Notes}},
\href{http://arxiv.org/abs/1812.01491}{{\ttfamily arXiv:1812.01491 [hep-ph]}}.

\bibitem{Borowka:2017idc}
S.~Borowka, G.~Heinrich, S.~Jahn, S.~P. Jones, M.~Kerner, J.~Schlenk, and
  T.~Zirke, {\it {pySecDec: a toolbox for the numerical evaluation of
  multi-scale integrals}},
  \href{http://dx.doi.org/10.1016/j.cpc.2017.09.015}{{\em Comput. Phys.
  Commun.} {\bfseries 222} (2018) 313--326},
\href{http://arxiv.org/abs/1703.09692}{{\ttfamily arXiv:1703.09692 [hep-ph]}}.

\bibitem{vonManteuffel:2014qoa}
A.~von Manteuffel, E.~Panzer, and R.~M. Schabinger, {\it {A quasi-finite basis
  for multi-loop Feynman integrals}},
  \href{http://dx.doi.org/10.1007/JHEP02(2015)120}{{\em JHEP} {\bfseries 02}
  (2015) 120},
\href{http://arxiv.org/abs/1411.7392}{{\ttfamily arXiv:1411.7392 [hep-ph]}}.

\bibitem{Smirnov:2020quc}
A.~Smirnov and V.~Smirnov, {\it {How to choose master integrals}},
  \href{http://arxiv.org/abs/2002.08042}{{\ttfamily arXiv:2002.08042
  [hep-ph]}}.

\bibitem{Usovitsch:2020jrk}
J.~Usovitsch, {\it {Factorization of denominators in integration-by-parts
  reductions}},  \href{http://arxiv.org/abs/2002.08173}{{\ttfamily
  arXiv:2002.08173 [hep-ph]}}.

\bibitem{Hahn:2004fe}
T.~Hahn, {\it {CUBA: A Library for multidimensional numerical integration}},
  \href{http://dx.doi.org/10.1016/j.cpc.2005.01.010}{{\em Comput. Phys.
  Commun.} {\bfseries 168} (2005) 78--95},
\href{http://arxiv.org/abs/hep-ph/0404043}{{\ttfamily arXiv:hep-ph/0404043
  [hep-ph]}}.

\bibitem{Ellis:2016jkw}
J.~Ellis, {\it {TikZ-Feynman: Feynman diagrams with TikZ}},
  \href{http://dx.doi.org/10.1016/j.cpc.2016.08.019}{{\em Comput.\ Phys.\
  Commun.} {\bfseries 210} (2017) 103--123},
  \href{http://arxiv.org/abs/1601.05437}{{\ttfamily arXiv:1601.05437
  [hep-ph]}}.

\bibitem{nicolas_deutschmann_2016_164393}
N.~Deutschmann, {\it {ndeutschmann/qgraf-xml-drawer: Qgraf-XML-drawer 1.0}},
  Nov., 2016.
\newblock \url{https://doi.org/10.5281/zenodo.164393}.

\bibitem{Nogueira:1991ex}
P.~Nogueira, {\it {Automatic Feynman graph generation}},
\href{http://dx.doi.org/10.1006/jcph.1993.1074}{{\em J. Comput. Phys.}
  {\bfseries 105} (1993) 279--289}.

\bibitem{Vermaseren:2000nd}
J.~A.~M. Vermaseren, {\it {New features of FORM}},
\href{http://arxiv.org/abs/math-ph/0010025}{{\ttfamily arXiv:math-ph/0010025
  [math-ph]}}.

\bibitem{Lee:2012cn}
R.~N. Lee, {\it {Presenting LiteRed: a tool for the Loop InTEgrals REDuction}},
\href{http://arxiv.org/abs/1212.2685}{{\ttfamily arXiv:1212.2685 [hep-ph]}}.

\bibitem{Studerus:2009ye}
C.~Studerus, {\it {Reduze-Feynman Integral Reduction in C++}},
  \href{http://dx.doi.org/10.1016/j.cpc.2010.03.012}{{\em Comput. Phys.
  Commun.} {\bfseries 181} (2010) 1293--1300},
\href{http://arxiv.org/abs/0912.2546}{{\ttfamily arXiv:0912.2546
  [physics.comp-ph]}}.

\bibitem{vonManteuffel:2012np}
A.~von Manteuffel and C.~Studerus, {\it {Reduze 2 - Distributed Feynman
  Integral Reduction}},
\href{http://arxiv.org/abs/1201.4330}{{\ttfamily arXiv:1201.4330 [hep-ph]}}.

\bibitem{fermat}
R.~H. Lewis, {\it {Computer Algebra System Fermat}},
  \href{http://dx.doi.org/http://home.bway.net/lewis/}{{\em
  http://home.bway.net/lewis/} }.

\end{thebibliography}\endgroup
\bibliographystyle{utphysM}
\end{document}